\documentclass[twocolumn]{aastex61}

\usepackage{natbib}
\usepackage{epstopdf}
\usepackage{subfigure}
\usepackage{amsmath}
\usepackage{todonotes}
\usepackage{color}

\usepackage{times}

\usepackage{longtable}

\shorttitle{Alfv\'enic Wave Heating}
\shortauthors{Reep et al.}

\begin{document}

\title{A Hydrodynamic Model of Alfv\'enic Wave Heating in a Coronal Loop and its Chromospheric Footpoints}

\author[0000-0003-4739-1152]{Jeffrey W. Reep}
\affiliation{National Research Council Postdoctoral Fellow, Space Science Division, Naval Research Laboratory, Washington, DC 20375, USA; \href{mailto:jeffrey.reep.ctr@nrl.navy.mil}{jeffrey.reep.ctr@nrl.navy.mil}}
\author{Alexander J.B. Russell}
\affiliation{School of Science \& Engineering, University of Dundee, Dundee DD1 4HN, UK}
\author{Lucas A. Tarr}
\affiliation{National Research Council Postdoctoral Fellow, Space Science Division, Naval Research Laboratory, Washington, DC 20375, USA}
\affiliation{Now at George Mason University}
\author{James E. Leake}
\affiliation{NASA Goddard Space Flight Center, Greenbelt, MD 20771, USA}

\begin{abstract}
Alfv\'enic waves have been proposed as an important energy transport mechanism in coronal loops, capable of delivering energy to both the corona and chromosphere and giving rise to many observed features, of flaring and quiescent regions.  In previous work, we established that resistive dissipation of waves (ambipolar diffusion) can drive strong chromospheric heating and evaporation, capable of producing flaring signatures.  However, that model was based on a simplified assumption that the waves propagate instantly to the chromosphere, an assumption which the current work removes.  Via a ray tracing method, we have implemented traveling waves in a field-aligned hydrodynamic simulation that dissipate locally as they propagate along the field line.  We compare this method to and validate against the magnetohydrodynamics code Lare3D.  We then examine the importance of travel times to the dynamics of the loop evolution, finding that (1) the ionization level of the plasma plays a critical role in determining the location and rate at which waves dissipate; (2) long duration waves effectively bore a hole into the chromosphere, allowing subsequent waves to penetrate deeper than previously expected, unlike an electron beam whose energy deposition rises in height as evaporation reduces the mean-free paths of the electrons; (3) the dissipation of these waves drives a pressure front that propagates to deeper depths, unlike energy deposition by an electron beam.
\end{abstract}


\section{Introduction}  
\label{sec:intro}

Alfv\'en waves, first predicted by \citet{alfven1942} and verified by \citet{lundquist1949a,lundquist1949b}, occur throughout the solar atmosphere, from the chromosphere \citep{depontieu2007} through the corona \citep{tomczyk2007, mcintosh2011}, extending out into the heliosphere with the solar wind \citep{belcherdavis1971}.  The dissipation of waves has been shown to heat and accelerate particles in solar and stellar winds \citep{suzuki2005,vanballegooijen2016} as well as cosmic rays \citep{fermi1949,lazarian2016}.  Their potential to heat the corona was quickly recognized \citep{alfven1947}, and heating due to their dissipation is still considered a plausible solution to the coronal heating problem \citep{klimchuk2006,vanballegooijen2011,reale2014}.  Alfv\'en waves are generated in the convection zone, as well as in the chromosphere by shock collisions \citep{osterbrock1961}, where they propagate upwards, possibly dissipating their energy in the chromosphere \citep{arber2016}.  A ponderomotive force due to the propagation of Alfv\'en waves acting on ions in the chromosphere has been suggested as an explanation for the first ionization potential (FIP) effect \citep{laming2015}, as well as affecting the streaming of stellar winds \citep{belcher1971}, and the force due to downward waves can drive acoustic waves that can then cause sunquakes \citep{russell2016}.  The recent article by \citet{russell2017} summarizes some of the major developments in the 75-year history of Alfv\'en waves in solar physics.

In this work, we focus on Alfv\'en waves generated via reconnection events in the corona.  Although it is expected that waves are generated during the reconnection events that drive flares and nanoflares \citep{parker1991,takeuchi2001,shibata2003,kigure2010,jelinek2017}, it is not currently known what fraction of the released energy they carry, with what frequencies they oscillate, or whether they cause heating that helps to power the emitted radiation.  Simple arcade models suggest that the waves occur across a spectrum without an upper bound on frequency \citep{oliver1993,tarr2017}.  \citet{fletcher2008} showed that Alfv\'en waves generated in the corona and propagating downwards might carry a significant fraction of the released magnetic energy during a flare, later supported by simulations of three-dimensional reconnection \citep{birn2009}, and they perhaps could cause acceleration of particles in the chromosphere.  \citet{haerendel2012} examined the possibility that the auroral acceleration process may take place in solar flares, whereby a release of magnetic shear stress due to Alfv\'en waves is converted into kinetic energy of particles.  A recent study with the New Solar Telescope found evidence for downward moving waves in a large solar flare, where their impact in the lower atmosphere possibly triggered a sunspot rotation \citep{liu2016}.  

The propagation of waves into the chromosphere has received significant attention in recent years.  \citet{haerendel2009} showed that such waves carry sufficient energy to heat the chromosphere via turbulent phase mixing, and thus drive chromospheric evaporation.  With a linearized magnetohydrodynamics (MHD) model, \citet{russellstackhouse2013} confirmed that Alfv\'en waves can deliver concentrated energy flux to the chromosphere.  \citet{russellfletcher2013} developed a model of energy transmission as waves propagate beyond the transition region and found strong resistive damping in the chromosphere due to ion-neutral friction, with a strong dependence on frequency.  Ion-neutral friction is well established as one of the primary damping mechanisms of Alfv\'en waves \citep{piddington1956}, and in recent years the governing theory has received significant attention \citep{soler2013, leake2014, soler2015a, khomenko2017}.  The effectiveness of ion-neutral friction depends on the local field strength, and is most efficient when the Lorentz force is strong compared to other dynamic forces \citep{soler2015b}.

In partially ionized plasma, the components of an electric current parallel or perpendicular to the magnetic field dissipate differently, due to collisions between the ionized plasma and neutral gas.  This anisotropic dissipation has been extensively covered in a broad range of fields, such as ionospheric/thermospheric physics, astrophysics, as well as solar physics.  While the governing physics remain the same, the breadth of application has led to different names and conceptualizations being used to describe the dissipation of perpendicular currents by plasma-neutral collisions: Cowling resistivity \citep{cowling1956,leake2005}, Pedersen resistivity \citep{haerendel2006,goodman2011}, and ambipolar diffusion \citep{zweibel1989,martinezsykora2012,martinezsykora2017} are all terms related to the dissipation of perpendicular currents by plasma-neutral collisions.  The discussions by \citet{zweibel2011} and \citet{leake2014} explain the differences of origin and emphasis between the terms. The recent applications in solar physics include wave propagation and dissipation \citep{depontieu1999,depontieu2001,leake2005,Soler2016,Brady2016}, mass and magnetic flux transport \citep{Leake2013, martinezsykora2012,martinezsykora2017}, chromospheric heating \citep{Goodman2010}, and magnetic reconnection \citep{Ni2015}.   Applied to our topic, the various terms are simply different names for the same anisotropic dissipation of electric currents.  We will primarily use the terminology of ion-neutral friction or Cowling resistivity.

A few models have been developed to study the hydrodynamic evolution of solar plasma due to wave heating.  \citet{emslie1982} developed a WKB formulation of resistive dissipation of Alfv\'enic waves launched in the corona in order to explain temperature minimum heating observed in large flares \citep{machado1978,emslie1979}.  Adopting this model with a correction for ambipolar diffusion, \citet{reep2016} showed that waves can also strongly heat the upper chromosphere and produce explosive evaporation.  \citet{kerr2016} confirmed these results, and further showed that chromospheric lines such as \ion{Mg}{2} could distinguish between Alfv\'enic wave heating and electron beam heating in flares, offering an important observational test.  In these studies, however, it was assumed that the waves propagate instantly from their injection location to the depths where they damp.  The assumption is justified for heating by a beam of electrons, where electrons with energy $> 10$\,keV travel at speeds $\gtrsim 0.2c \approx 60$\,Mm\,s$^{-1}$.  Waves, however, travel at the local Alfv\'en speed, which is typically $< 10$\,Mm\,s$^{-1}$ in the corona (although it may approach $0.1c$ in the corona of some active regions \citealt{fletcher2008,russellfletcher2013}), and significantly slower elsewhere.  The damping of waves affects the atmosphere on a similar time-scale as the wave travel times, so travel times should not be neglected.  

The model in \citet{reep2016} required high frequencies ($\gtrsim 1$\,Hz) for effective dissipation of wave energy, and so the frequency spectrum of waves in the corona is important to study.  \citet{deforest2004} directly observed frequencies as high as 0.1\,Hz with TRACE in the 1600\,\AA\ passband.  Eclipse observations have also shown the presence of intensity oscillations of the order 1-10\,Hz \citep{pasachoff2002,rudawy2004}, although it is possible that those fluctuations are not due to Alfv\'en waves \citep{rudawy2010}.  Microwave and radio bursts during solar flares have revealed oscillations of the order 10--100\,Hz \citep{kiplinger1983,kaufmann1984}.  As these frequencies are at the limit of current instrumental cadence, it is unclear whether higher cadence instruments would reveal the presence of higher frequencies, and if so, their importance to the physical processes occurring in the solar atmosphere.  The next generation of solar instrumentation, in particular the suite of instruments to come online in 2019--2020 at the Daniel K. Inouye Solar Telescope (DKIST,  \citealt{elmore2014}), will have the cadence and sensitivity to probe the super-hertz frequency range of MHD waves in the low corona, offering a first look at whether such waves are present in the corona, if they are generated by reconnection, and what their energetic importance may be.

In this work, we directly examine the propagation of Alfv\'en waves by implementing a ray tracing code that follows the waves as they travel and damp along a coronal loop in a one-dimensional hydrodynamics simulation.  We describe the implementation and necessary physics in Section \ref{sec:implementation}.  We compare this implementation to the MHD code Lare3D in Section \ref{sec:larecomparison}.  We then directly compare the results of heating due to propagating waves with the previous work that assumed instantaneous travel times, as well as with an electron beam, in Section \ref{sec:results}.

\section{Relevant Physics \& Implementation}  
\label{sec:implementation}

We have implemented traveling Alfv\'en waves in the HYDrodynamics and RADiation code (HYDRAD, \citealt{bradshaw2003}), which solves the equations of conservation of mass, momentum, and energy for a two-fluid plasma confined to an isolated magnetic flux tube (equations and details are listed in \citealt{bradshaw2013}).  The code includes adaptive mesh refinement of arbitrary order, important for resolving rapid jumps in the temperature or density profiles, which can lead to improper estimates of coronal properties if not properly treated \citep{bradshaw2013}.  Radiative losses are treated with a full calculation of emissivities from abundant ions using CHIANTI version $8$ \citep{dere1997,delzanna2015}, with the ability to simultaneously solve for non-equilibrium ionization populations of any desired element \citep{bradshaw2013b}.  The code is quick, robust, and computationally inexpensive.  

Alfv\'en waves have long been considered important to the dynamics of energy transport in the solar atmosphere.  They cannot be treated in pure hydrodynamic (HD) codes, which are designed to study energy transport but do not solve the full set of MHD equations.  On the other hand, MHD codes generally do not have a full treatment of thermodynamics, or the proper resolution to resolve the transition region, which often means it is difficult to properly understand the evolution of a loop's hydrodynamics \citep{bradshaw2013}.  We wish to bridge the gap between these two approaches, and implement a method that allows us to study Alfv\'en wave dissipation in a HD code.

\begin{figure}
\begin{minipage}[b]{\linewidth}
\centering
\includegraphics[width=\textwidth]{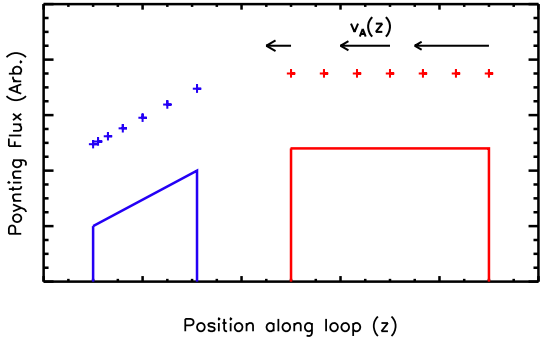}
\end{minipage}
\caption{Example diagram of a pulse traveling to the left (solid, represented as a Poynting flux - energy per area per time), being traced with seven rays in time (plus signs to denote locations).  Each ray travels at the local Alfv\'en speed, which can vary in both time and position, assumed to be decreasing to the left here.  As the rays propagate, the Poynting flux at each ray decreases as the pulse damps, but they become ``bunched up'' in space.  In general, the shape of the Poynting flux curve with time depends on the variable spacing between rays and the flux at each one.  }
\label{fig:waveschematic}
\end{figure}

We therefore introduce a new method to describe waves as pulses of period-averaged Poynting flux propagating along a loop, which we trace with individual rays at various points along the pulse.  Figure \ref{fig:waveschematic} illustrates this.  We inject a pulse with a certain initial Poynting flux, comprised of a number of rays $N$ that approximates the spatial extent of a pulse.  In Figure \ref{fig:waveschematic}, this initial pulse is represented by the solid red curve, which has been divided into 7 rays, each with their own position and Poynting flux (red plus signs, offset by an arbitrary factor).  In order to trace the motion of the whole pulse, each ray is individually advected along the loop in the direction of propagation at the local Alfv\'en speed (which may differ for each ray), and the Poynting flux is decreased according to the local damping length (see Section \ref{subsec:raytracing}).  The blue curve in Figure \ref{fig:waveschematic} represents the pulse at some time later, where the Poynting flux has decreased, and the spatial extent has narrowed due to a decreasing Alfv\'en speed.  As a result, the rays are closer together, especially near the leading edge of the pulse.

In order to inject a pulse onto the loop, we initialize the rays one at a time.  If the pulse is injected at time $t_{0}$, the first ray is created at time $t_{0}$, with a position $z_{0}$ corresponding to the location of injection and Poynting flux equal to the pulse's initial flux $S_{0}$.  Supposing that there are $N > 1$ total rays comprising the pulse, and the duration of the pulse is $\tau_{\text{dur}}$, the next ray is created at time $t_{1} = t_{0} + \tau_{\text{dur}} / (N - 1)$, the next at $t_{2} = t_{1} + \tau_{\text{dur}} / (N - 1)$, and so on until the last ray is created at time $t_{N-1} = t_{0} + \tau_{\text{dur}}$.  Between times $t_{0}$ and $t_{1}$, the first ray is appropriately advected, between $t_{1}$ and $t_{2}$, the first two rays are advected, {\it etc}.  

The number of rays $N$ comprising a pulse must be high enough to both capture the spatial extent of that pulse and to capture small scale changes in the Alfv\'en speed and damping lengths, particularly in regions of large gradients.  If the duration is longer, the spatial extent is greater, and so requires more rays to capture the full extent and sharp gradients.  Further, with the WKB method, we need not resolve the wavelength, but we do need to resolve changes in the damping length.  We therefore choose the number of rays $N$ to be 50 rays per second of injection duration, or $N = 50\ \tau_{\text{dur}}$, which we have found through simple tests to accurately capture the heating while not being too demanding computationally.  

The input for the waves is general: the user specifies the number of pulses, and, for each one, the start time, duration, injection location along the loop, direction of propagation, initial Poynting flux, frequency, and perpendicular wave number (at the photosphere).  The code then divides each pulse up into a number of rays that are traced in time and position.  

\subsection{Ray Tracing}
\label{subsec:raytracing}

At each numerical time step, the code first checks to see if any new rays need to be initialized, {\it i.e.} if the current time is between the starting injection time $t_{0}$ and the final injection time $t_{0} + \tau_{\text{dur}}$ for any pulse.  If so, new rays are created at the specified location with the specified properties of the pulse to which they belong (as described in the previous section).  

All currently existing rays are then advected at each time step.  The rays travel at the local Alfv\'en speed $v_{A}(z)$, from an initial position $z_{0}$ to a final position $z_{f}$.  In general, the pulse will have arbitrary spatial extent and span numerous grid cells, each with a distinct Alfv\'en speed, so that the rays propagate at different speeds.  The ray positions are continuous, while the physical variables (and the local Alfv\'en speed defined by them) are defined on the discrete grid.  We must therefore carefully treat each ray's propagation and damping within the discrete system.  

Suppose the bounds of a grid cell are given by $z_{1}$ and $z_{2}$, with Alfv\'en speed $v_{A}$ between those bounds.  We assume that the Alfv\'en speed is constant within a grid cell for simplicity, though a more general treatment could interpolate the value for any point within the cell.  For a ray position $z_{1} < z_{\text{ray}} < z_{2}$, the time to traverse the grid cell is given by 
\begin{align}
\tau = \frac{z_{2} - z_{\text{ray}}}{v_{A}}  
\end{align}
\noindent We have two cases then.  In the first, the numerical time step $\Delta t < \tau$, {\it i.e.} that the current time step is smaller than the time for the ray to move to the next grid cell, in which case the ray is advected by
\begin{align}
\Delta z = v_{A} \Delta t
\label{eqn:simpleadvect}
\end{align}
\noindent In the second case, $\Delta t > \tau$, {\it i.e.} that the ray moves into a new grid cell in the current time step, we use a recursive function to then find the new ray position.  Numbering each successive grid cell from the starting position of the ray, the new ray position is then:
\begin{align}
\Delta z &= v_{A, 0}\ \tau_{0} + v_{A, 1}\ \tau_{1} + ... + v_{A, N}\ \tau_{N} \\ \nonumber
	&= \sum_{i = 0}^{N} v_{A, i}\ \tau_{i}
\end{align}
\noindent which is continued recursively while $\Delta t > \sum_{i = 0}^{N} \tau_{i}$.  When that condition fails, we advance the ray in the final grid cell by the distance $v_{A} \times (\Delta t - \sum_{i = 0}^{N} \tau_{i})$.    

All of the rays are advected appropriately at each time step with this method.  Simultaneously, we calculate the reduction in their Poynting fluxes due to resistive dissipation according to the local damping lengths (given in Section \ref{subsec:heating}).  Specifically, the decrease of Poynting flux from $z_{0}$ to $z_{f}$ follows the general equation
\begin{align}
S(z_{f}) = S(z_{0}) \exp{\Bigg(- \int_{z_{0}}^{z_{f}} \frac{dz^{'}}{L_{D}(z^{'})}\Bigg)}
\end{align}
\noindent which is applied to each ray as their locations are updated.

\subsection{Heating}
\label{subsec:heating}

The heating terms follow the same basic equations as in \citet{reep2016}, except that the heated locations are now limited to grid cells which contain a pulse at a given time.  The code loops across the numerical grid, locating the start and end positions of each pulse ({\it i.e.} the positions of the first and last ray in a pulse).  If a grid cell is located between the leading and trailing edge of a pulse, then the grid cell is assumed to be heated by the resistive dissipation of that pulse as it passes through.  After the rays have been advected forward and their Poynting fluxes appropriately decreased, that lost energy is taken to heat the local plasma.  The heating function is given in general by:
\begin{align}
Q(z) &= - \frac{d S(z)}{dz} \\ \nonumber
	&= \frac{S(z)}{L_{D}(z)} \\ \nonumber
	&= S(z) \Big( \frac{1}{L_{e^{-}}} + \frac{1}{L_{H}}\Big) \\ \nonumber
	&\equiv Q_{e^{-}} + Q_{H}
\label{eqn:heating}
\end{align}
\noindent which defines the split between the electron heating term and the hydrogen heating term in terms of the damping lengths:
\begin{align} 
Q_{e^{-}} &= \frac{S(z)}{L_{e^{-}}} \\
Q_{H} &= \frac{S(z)}{L_{H}} 
\end{align}
\noindent The damping lengths are given by
\begin{align}
\frac{1}{L_{D}} &= \frac{1}{L_{e^{-}}} + \frac{1}{L_{H}}  \\ \nonumber
	&= \Bigg[ \eta_{\parallel} \Big(\frac{k_{x}^{2}c^{2}}{4\pi v_{A}} + \frac{\omega^{2}c^{2}}{4\pi v_{A}^{3}}\Big) \Bigg] + \Bigg[ \eta_{C} \Big( \frac{\omega^{2}c^{2}}{4\pi v_{A}^{3}}\Big) \Bigg]
\label{eqn:dampinglengths}
\end{align}
\noindent where $k_{x}(z)$ is the local perpendicular wave number (which we assume varies as $k_{x}(z) = k_{x,a} \frac{B}{B_{a}}$ for $k_{x,a}$ and $B_{a}$ the values at the apex of the loop), $\omega$ the angular frequency, $c$ the speed of light in vacuum, $v_{A}(z)$ the local Alfv\'en speed, and $\eta_{\parallel}(z)$ and $\eta_{C}(z)$ the parallel and Cowling resistivities (due to ion-neutral friction, \citealt{cowling1956}), given by  
\begin{align}
\eta_{\perp} &= \eta_{\parallel} + \eta_{C} \\ \nonumber
	&= \frac{m_{e}\ (\nu_{ei} + \nu_{en})}{n_{e}e^{2}} + \frac{B^{2} \rho_{n}}{c^{2} \nu_{ni} \rho_{t}^{2}\ (1 + \xi^{2} \theta^{2})}
\end{align}
\noindent where $m_{e}$ is the electron mass, $B(z)$ the magnetic field strength, $\rho(z)$ the mass density, and $\nu(z)$ the collision frequencies.  The subscripts $i$, $e$, $n$, $t$ refer to electron, ion, neutral, and total, respectively.  As in \citet{reep2016}, we define $\xi = \frac{\rho_{i}}{\rho_{t}}$ and $\theta = \frac{\omega}{\nu_{ni}}$, and modify the Alfv\'en speed due to the presence of neutrals:
\begin{align}
v_{A}(z) = \frac{B}{\sqrt{4 \pi \rho_{t}}} \Bigg( \frac{1 + \xi \theta^{2}}{1 + \xi^{2} \theta^{2}} \Bigg)^{1/2}
\end{align}
\noindent This expression reduces to the standard form of the Alfv\'en speed in the limit of a fully ionized plasma.

The damping lengths are calculated locally at each time step, and each ray has its Poynting flux decreased using the local length in each cell it traverses.  The heating rate in a grid cell is then found by interpolating the Poynting flux at the center of the grid cell using the two nearest rays.  Consequently, there is a small error associated with the finite width of the grid cells: the heating rate is more accurate for small grid cells.  

In order to partition the energy between the two species, we can also rewrite the equations more directly in terms of the total heat:
\begin{align}
Q_{e^{-}}(z) = Q(z) \frac{L_{D}(z)}{L_{e^{-}}(z)}  \\
Q_{H}(z) = Q(z) \frac{L_{D}(z)}{L_{H}(z) } 
\end{align}
\noindent Then, each term is added to its respective energy equation.

\subsection{Time Step} 

As mentioned before, if the current time is between the initial injection time $t_{0}$ and final injection time $t_{0} + \tau_{\text{dur}}$ for any pulse, new rays must be created at an appropriate time step $\tau_{dur}/(N-1)$ for the number of rays $N$ in the pulse.  If $N$ is large, then this time step becomes tiny and may become smaller than the numerical time step $\Delta t$ of the code.  When this happens, some rays might fail to be properly injected.  To avoid this problem, we introduce a new time-scale to HYDRAD to insure that all rays are properly created.

At a given time, if there is at least one pulse being initialized, we set a new time-scale to one tenth of the time between each successive ray (which is rather cautious, for computational speed this could be increased up to one half of the time between rays).  That is, for a duration of injection $\tau_{\text{dur}}$ and number of rays $N$, the new time-scale is $\tau_{\text{rays}} = \frac{\tau_{\text{dur}}}{10 N}$.  Then, this time-scale is guaranteed to be smaller than the time between the creation of each ray, and each one will be properly initialized.  

This time-scale is then calculated for each pulse currently being injected, compared to the other relevant time-scales, and then the time step in the simulation is set to the minimum time-scale, as done in general with HYDRAD \citep{bradshaw2013}.  In this way, we insure that all rays are properly injected, and that the CFL condition is met.

\section{Comparison to Lare}  
\label{sec:larecomparison}

To validate that the ray tracing method reproduces the correct propagation and dissipation, we first compare our implementation of the WKB wave-packets with the magnetohydrodynamics Lagrangian-Eulerian Remap 3D code (Lare3D, \citealt{arber2001}).  We present brief results necessary for validation here, but postpone a detailed comparison of the HD and MHD codes for a later work.  In the next section, we present the detailed results of propagating wave simulations with HYDRAD.

Lare3D is a parallel code that solves the full set of non-linear MHD equations by first taking a Lagrangian step with each computational cell and, second, conservatively remapping the result back to the original Eulerian grid.  This method allows for shocks to be properly resolved and can be readily adapted to include most physical processes. We use the modified MHD equations which include the effects of partial ionization and Cowling resistivity \citep{leake2005}.  We use this modified Lare code in 1.5D (only one dependent variable, but all components of the 3D vector are evolved) to model a  one-dimensional coronal loop (similarly done by other authors, {\it e.g.} \citealt{johnston2017a, johnston2017b}), and introduce an Alfv\'enic wave pulse in order to compare to our new method in HYDRAD.  We use a grid spacing of $460$\,m in Lare, and do not include the effects of radiative losses or thermal conduction.  The ray tracing method presented here includes only Alfv\'enic perturbations.  Therefore in the Lare3D simulation that we use to verify and validate the method, we remove the non-linear coupling of Alfv\'enic perturbations into longitudinal perturbations.  Future studies will examine the fully non-linear MHD system, along with mode-coupling and multi-dimensions, as well as the thermodynamics appropriate for the coupled chromosphere-corona.

\begin{figure*}
\begin{minipage}[b]{\linewidth}
\centering
\includegraphics[width=\textwidth]{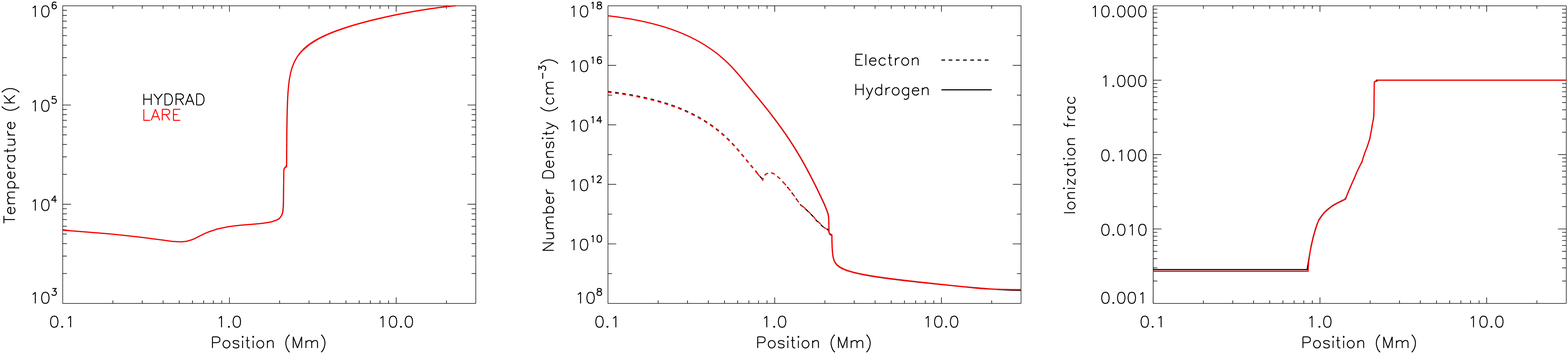}
\end{minipage}
\begin{minipage}[b]{\linewidth}
\centering
\includegraphics[width=\textwidth]{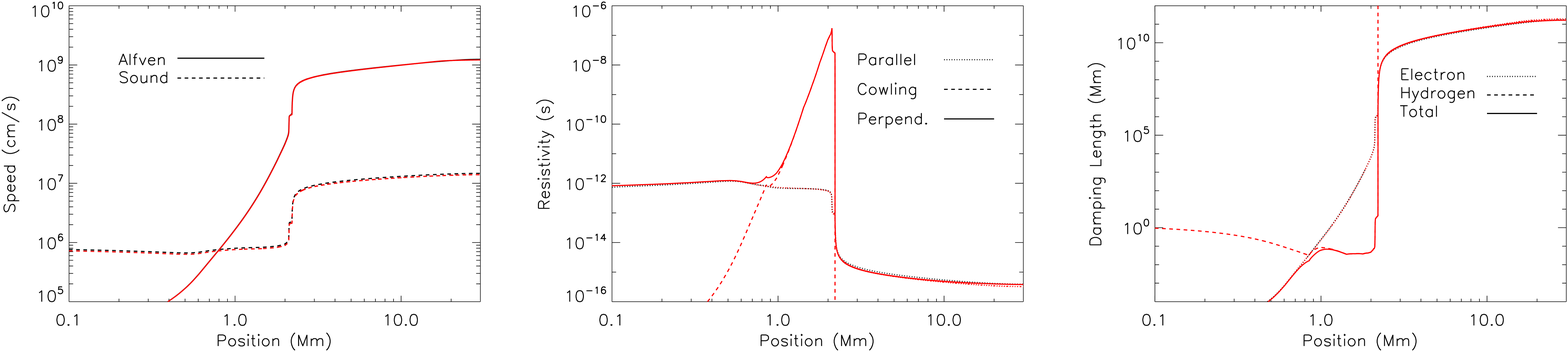}
\end{minipage}
\caption{A comparison of the initial values of the simulations for one half of a loop of length $2L = 60$\,Mm.  In all of the plots, HYDRAD values are shown in black, Lare in red.  At top, from left to right, the temperature, number densities, and ionization fraction of hydrogen as a function of position.  At bottom, for the assumed magnetic field strength profile, the Alfv\'en and sound speed, resistivities, and damping lengths of the waves (assuming frequency $f = 10$\,Hz, perpendicular wave number $k_{x} = 0$, and a constant $B = 107.9$\,G.  }
\label{fig:LC_initial}
\end{figure*}

We have tuned the initial conditions of the Lare simulation to match the HYDRAD ones as closely as possible.  We show the initial conditions of both codes in Figure \ref{fig:LC_initial}, from left to right and top to bottom: the temperature, number densities, ionization fraction, initial Alfv\'en speed and sound speed, resistivities, and damping lengths for both HYDRAD (black) and Lare (red), calculated with wave frequency $f = 10$\,Hz, perpendicular wave number $k_{x} = 0$, and $B(z)$ profile as in \citet{russellfletcher2013}.  Since $k_{x} = 0$, the current is perpendicular to the magnetic field and damping is determined by $\eta_{\perp}$.

\begin{figure}
\centering
\includegraphics[width=0.5\textwidth]{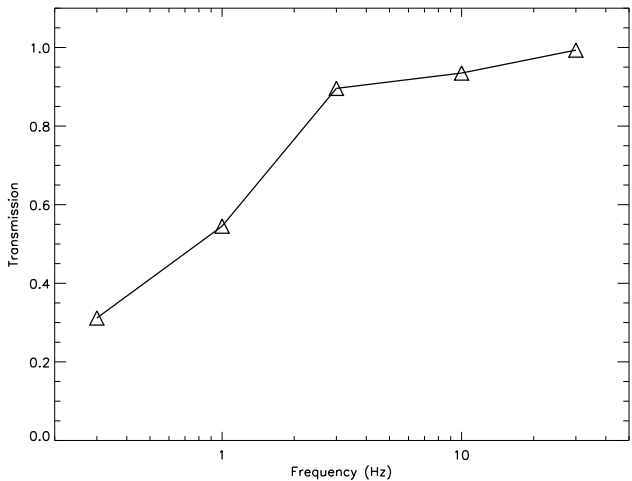}
\caption{Energy transmission coefficient at the transition region as a function of frequency in the Lare simuations for the chosen Alfv\'{e}n speed profile.  High frequency waves pass mostly unimpeded through the TR, while low frequency waves reflect a large portion of their Poynting flux.  The results are comparable with \citet{russellfletcher2013}.  
\label{fig:trans}}
\end{figure}

Although HYDRAD and Lare are given essentially the same initial conditions, they are fundamentally different codes that solve different equations, therefore the resulting dynamics will not always agree.  One major point of comparison is that we have not implemented reflection into HYDRAD, which arises in Lare because it solves the full MHD equations.  At the region of largest Alfv\'{en} speed gradient, around the transition region (Figure \ref{fig:LC_initial}), waves with low frequency will reflect back into the corona. Figure \ref{fig:trans} shows the energy transmission coefficient, measured as a ratio of the transmitted Poynting flux to the original Poynting flux, as a function of wave-packet frequency. While reflection at frequencies at or below 1\,Hz dominate, it is reasonable to assume that for frequencies $\gtrsim 2$ Hz, the use of the current HYDRAD implementation without reflections is reasonable, though the inclusion of reflection would be an important generalization of the current method.  

We have therefore run a simulation with both HYDRAD and Lare, using a frequency $f = 10$\,Hz and perpendicular wave number $k_{\text{x,a}} = 0$.  To partially account for the absence of reflection, we use a Poynting flux of $10^{9}$\,erg\,s$^{-1}$\,cm$^{-2}$ in Lare, which we reduce by the reflection coefficient ($\approx 10$\%) in HYDRAD.  In order to facilitate comparison, we do not allow the plasma to be heated, which would cause discrepancies in the temperature and ionization fraction between Lare and HYDRAD.  In other words, the initial profile is maintained throughout the simulation to ease the comparison.

\begin{figure*}
\begin{minipage}[b]{0.5\linewidth}
\centering
\includegraphics[width=\textwidth]{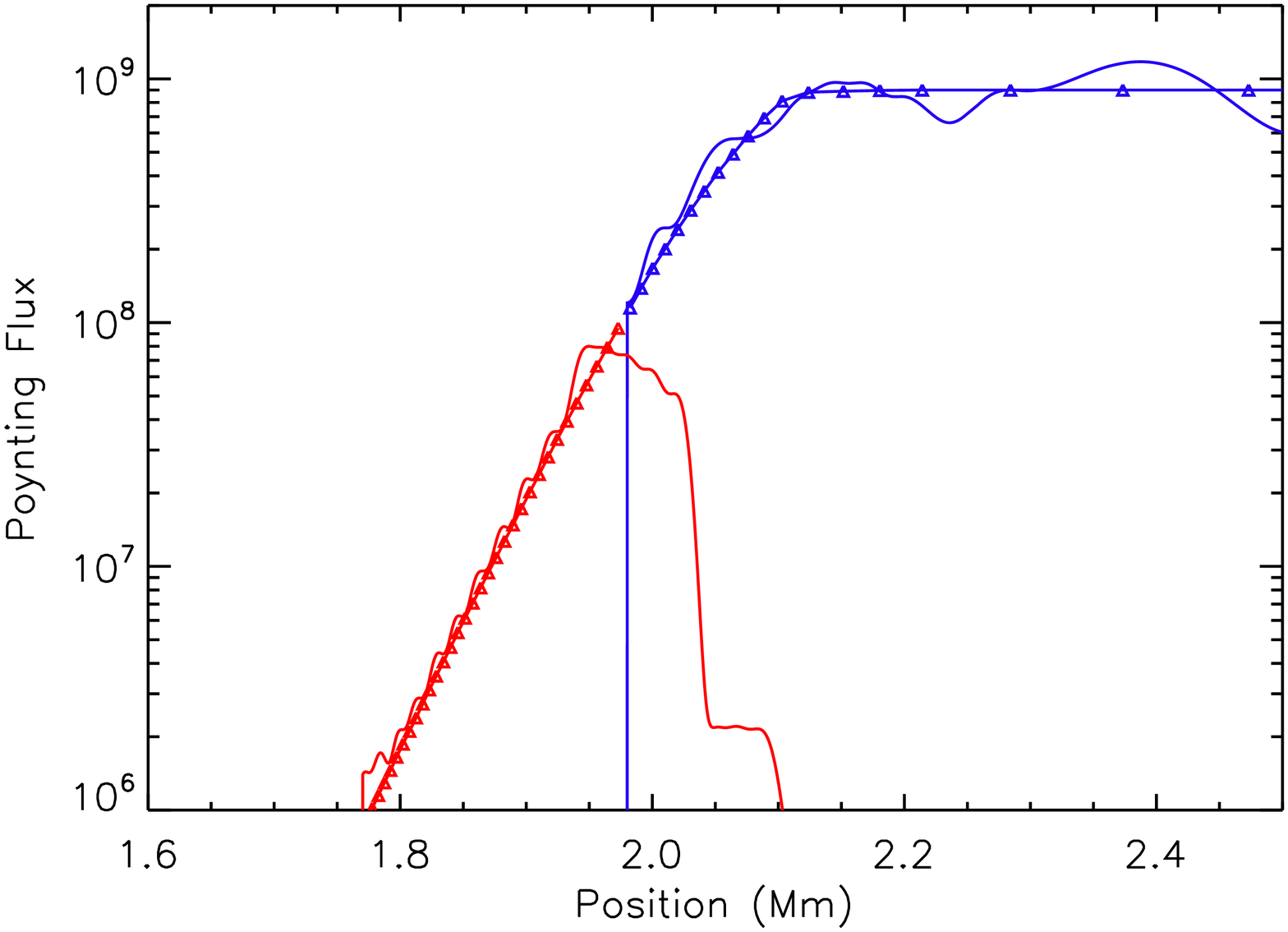}
\end{minipage}
\begin{minipage}[b]{0.5\linewidth}
\centering
\includegraphics[width=\textwidth]{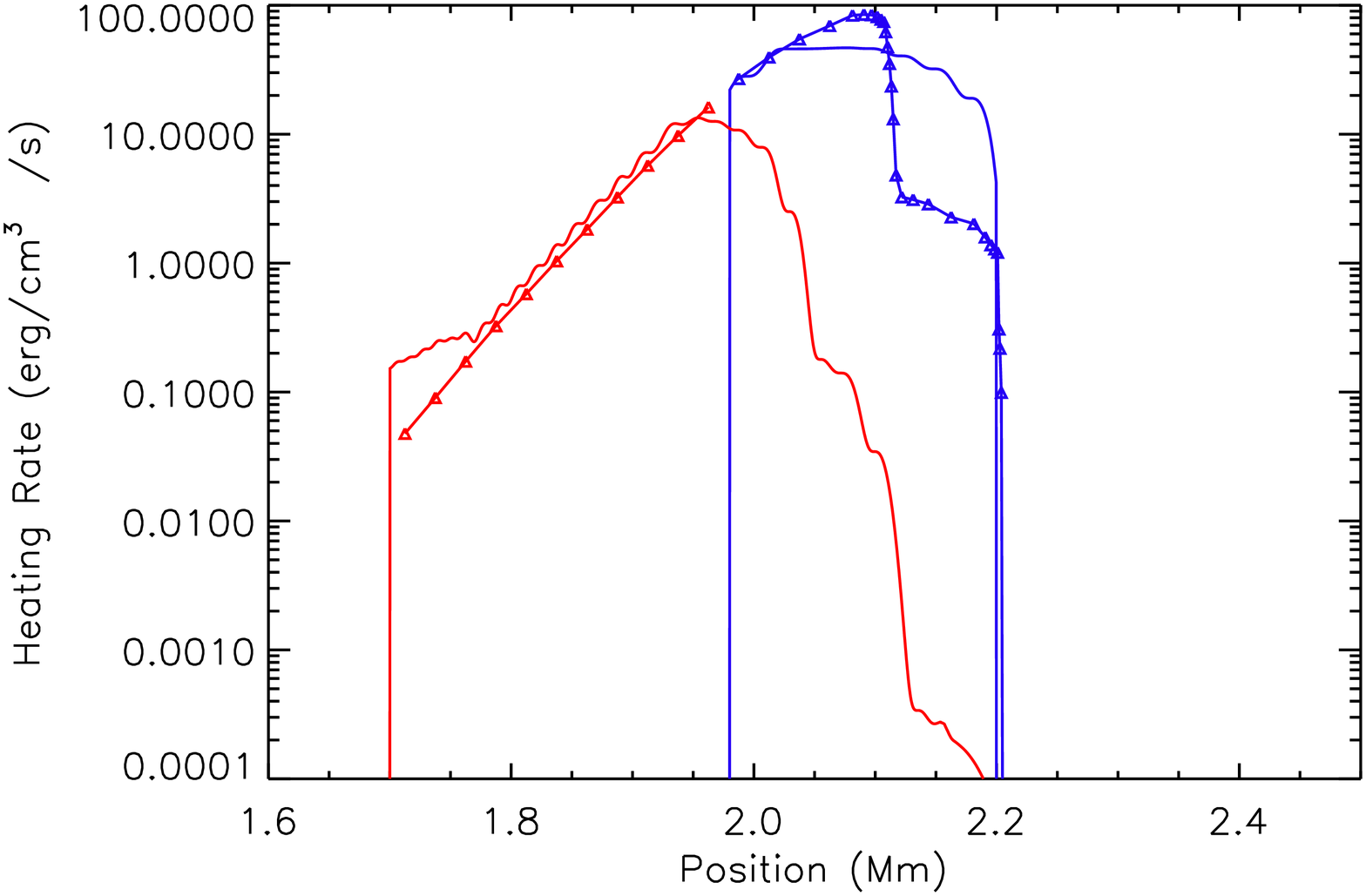}
\end{minipage}
\caption{A comparison between a wave simulation with HYDRAD (points) and Lare (lines), which compare favorably in terms of propagation, damping, and heating in general.  The waves have a frequency $f = 10$\,Hz, and initial Poynting flux of $10^{9}$\,erg\,s$^{-1}$\,cm${-2}$, reduced by about 10\% in the HYDRAD simulation to account for reflection.  The left hand plot shows the period-averaged Poynting flux at times 3\,s (blue) and 4\,s (red) into the simulation, shortly after the  pulse reaches the transition region where it strongly damps.  The right hand plot shows the heating rate derived from the damping at the same times.  }
\label{fig:LC_sim}
\end{figure*}

Figure \ref{fig:LC_sim} shows a few snapshots of this comparison simulation near the top of the chromosphere, shortly after the waves impinge upon it.  The plot on the left shows the period-averaged Poynting flux as a function of position, at times 3\,s (blue) and 4\,s (red) into the simulation.  HYDRAD is shown with points (each ray), and Lare as a solid line.  The plot on the right shows the derived heating rate as a function of position.  The slight discrepancies at the trailing edge of the pulse are due to reflection, where some of the Poynting flux has begun to propagate in the opposite direction in the Lare simulation.  

While there are differences between the two simulations, primarily due to the reflection which has not been included in HYDRAD, we find that the ray tracing method reproduces the approximate behavior of the MHD code.  There are differences that can be addressed in future iterations to further improve the comparison: the aforementioned reflection, the ponderomotive force, a method to treat mode conversion, {\it etc}.  Although these limitations exist, this method faithfully captures the propagation and resistive dissipation of waves through a field-aligned flux tube.  In the Appendix, in order to demonstrate the current limitations of this method, we show a comparison of a low-frequency case, where the effect of reflection is more pronounced.

\section{Results}
\label{sec:results}

The comparison to Lare validated our implementation of a WKB approximation for wave propagation and damping in a hydrodynamic code.  We now perform a detailed examination for different wave-packet parameters, with and without accounting for travel times.  We have run simulations with HYDRAD to examine the hydrodynamics.  We use a loop length $2L = 60$\,Mm, assumed semi-circular and oriented vertically relative to the solar surface.  We use the magnetic field strength profile $B(z)$ as specified in \citet{russellfletcher2013} in all the simulations here: $B(z) = B(0) \Big(\frac{P(z)}{P(0)}\Big)^{0.139}$ for the initial pressure profile calculated from the hydrostatic equations, and a constant field strength of $\approx 100$\,G in the corona.  We assume a photospheric value of 1000\,G, as in \citet{reep2016}.  This is an empirical fit intended to allow the field to vary smoothly across the chromosphere, though as \citet{russellfletcher2013} note, it is difficult to observationally measure the field strength across the chromosphere so that this is an assumed profile.  

\begin{figure*}
\begin{minipage}[b]{0.33\linewidth}
\centering
\includegraphics[width=\textwidth]{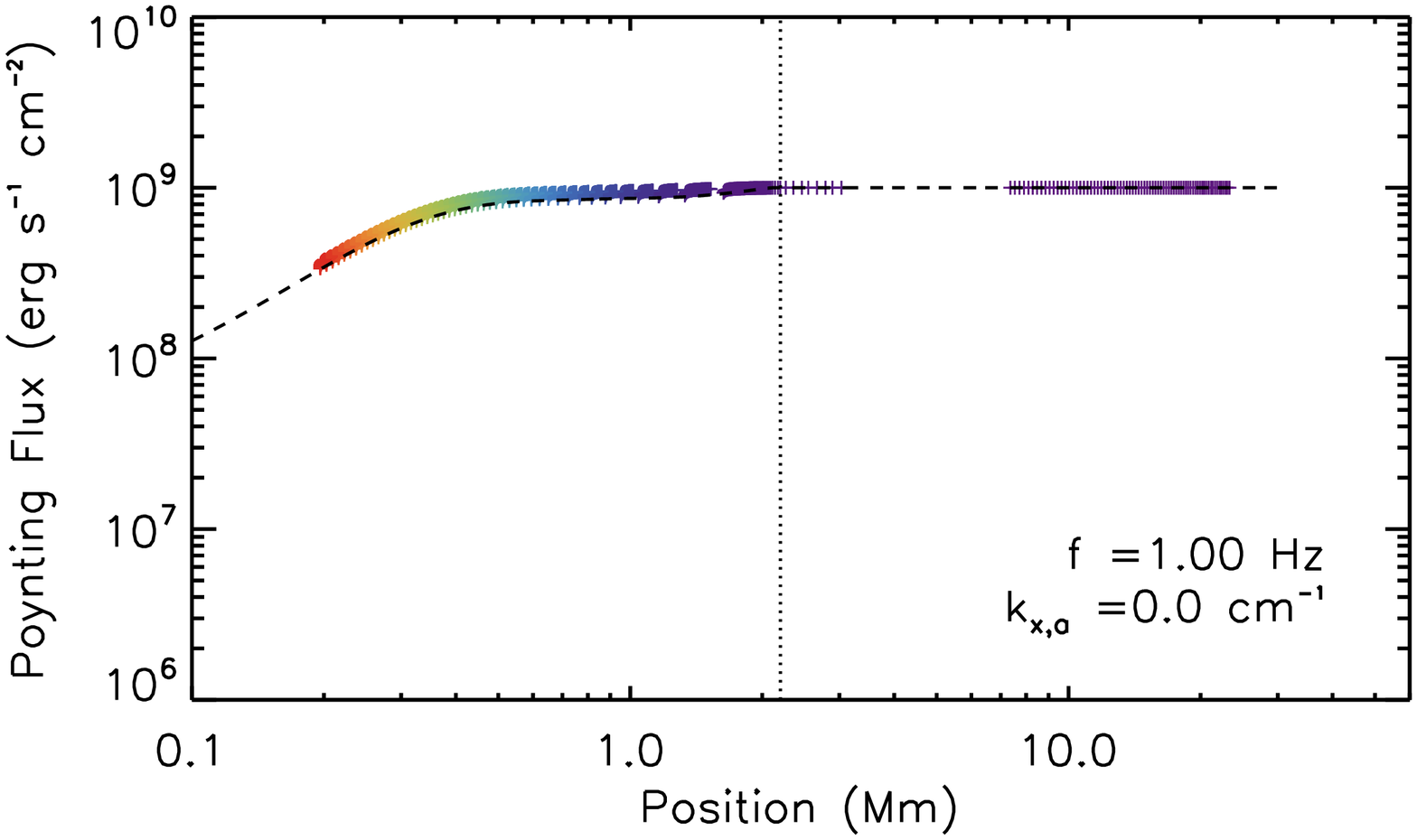}
\end{minipage}
\begin{minipage}[b]{0.33\linewidth}
\centering
\includegraphics[width=\textwidth]{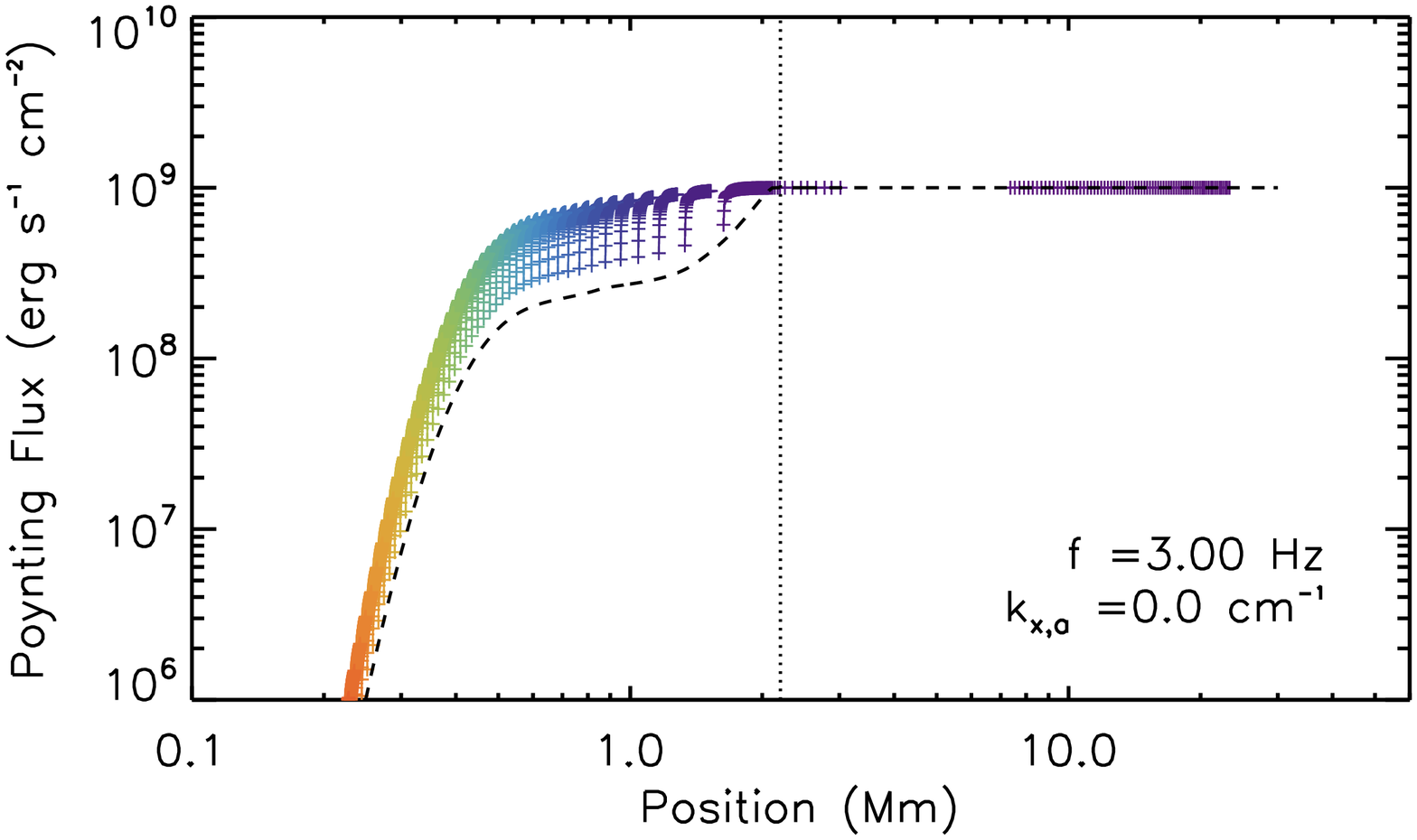}
\end{minipage}
\begin{minipage}[b]{0.33\linewidth}
\centering
\includegraphics[width=\textwidth]{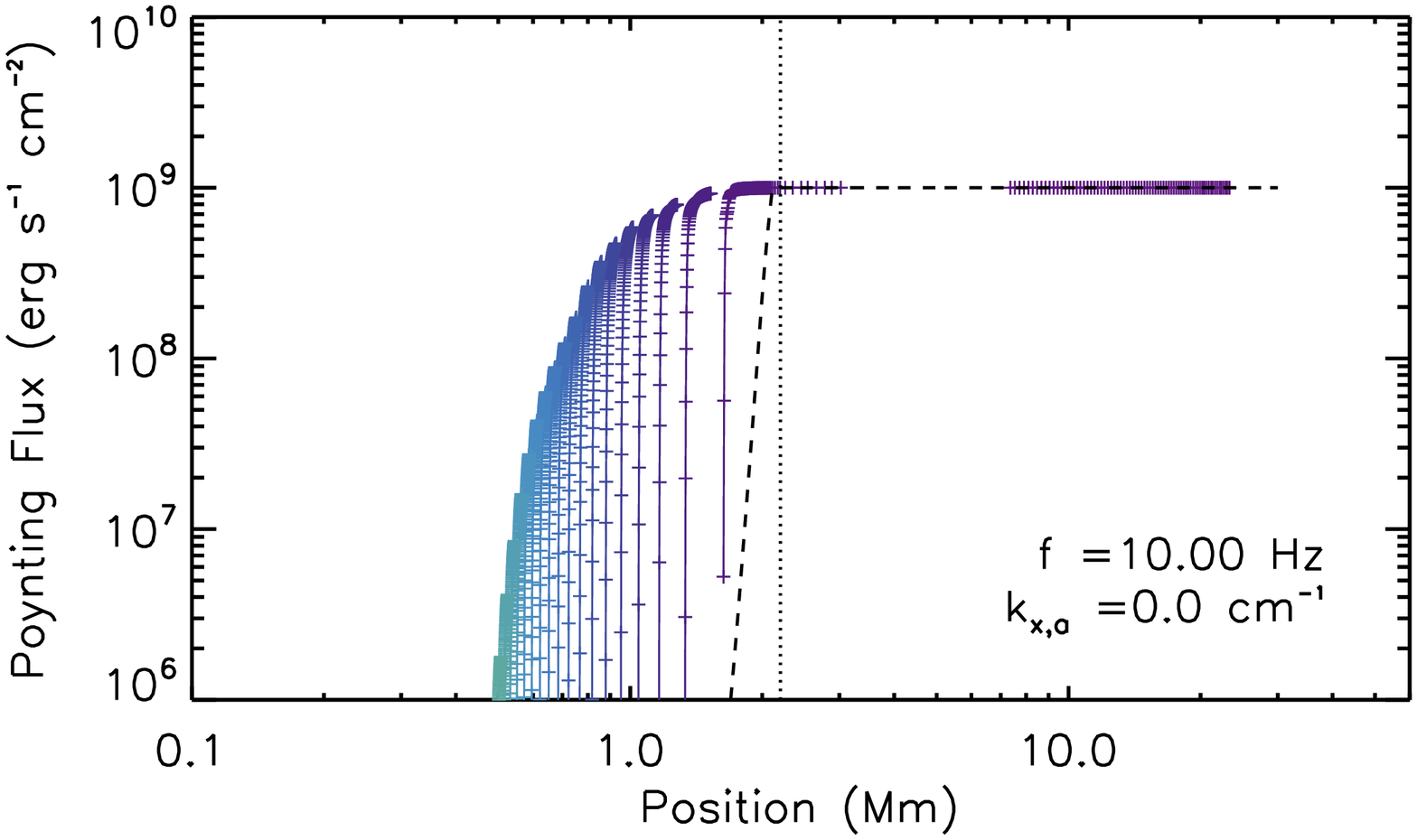}
\end{minipage}
\caption{The Poynting flux as a function of position and time (colors, from violet to red) for three simulations with a single 1-second pulse of frequency of 1 Hz (left), 3 Hz (center), and 10 Hz (right), and perpendicular wave number $k_{x} = 0$, traveling to the left-hand side of the loop.  Each plus sign denotes the location and flux of an individual ray, while the connecting solid lines denote the interpolated Poynting flux.  The times are plotted at a 2 second cadence, for a total of 100 seconds past the initial injection.  The black dashed line denotes the decrease of Poynting flux for the instant-travel method used in \citet{reep2016}, for comparison.     }
\label{fig:LC_poynting}
\end{figure*}

We first examine the damping of Poynting flux at various frequencies.  Figure \ref{fig:LC_poynting} shows the Poynting flux in three HYDRAD simulations for one-second pulses with $S_{0} = 10^{9}$\,erg\,s$^{-1}$\,cm$^{-2}$, $f = $[1, 3, 10]\,Hz, $k_{x} = 0$, propagating to the left.  The colors ranging from violet to red show different time steps, at a $2$\,s cadence, for a total of 100 seconds after the initial injection of the pulse.  Each plus sign denotes the location and flux of a single ray, while the lines connecting them denote the interpolated Poynting flux at a given position.  The pulses take just over 3 seconds to reach the transition region, where they begin to damp.  As each pulse propagates down the loop, the Poynting flux decreases appropriately, with high frequency pulses dissipating more rapidly, though even the 10\,Hz case takes well over 10 seconds to fall to 10\% of its initial Poynting flux.  By way of comparison, we over-plot the damping locations determined by the method used in \citet{reep2016}, shown as black dashed lines.  In all cases, the instant travel method damps slightly higher in the atmosphere than the ray tracing method, and the travel time is too significant to be ignored.  The difference in damping locations is exacerbated at high frequencies.  

\begin{figure*}
\begin{minipage}[b]{0.5\linewidth}
\centering
\includegraphics[width=\textwidth]{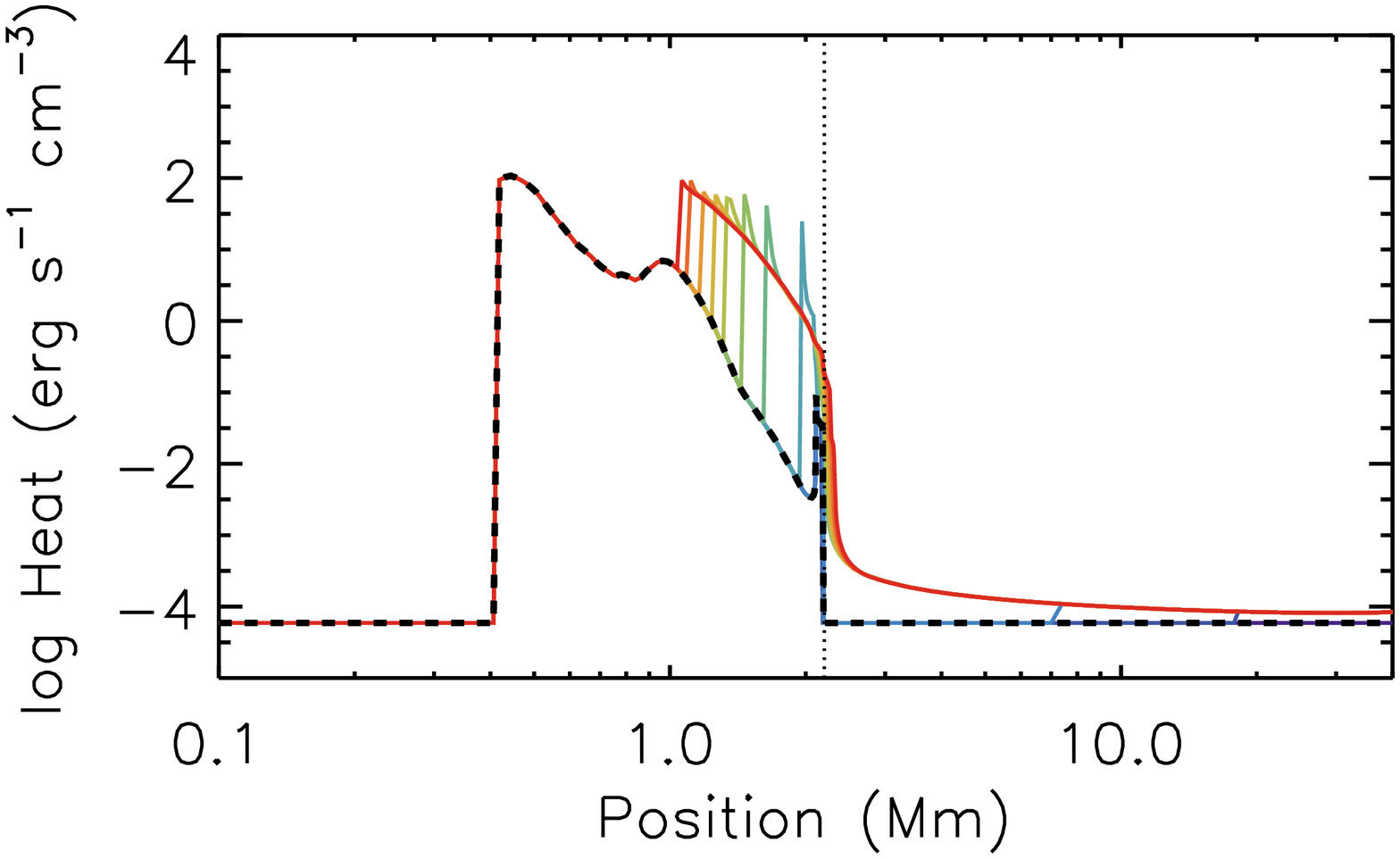}
\end{minipage}
\begin{minipage}[b]{0.5\linewidth}
\centering
\includegraphics[width=\textwidth]{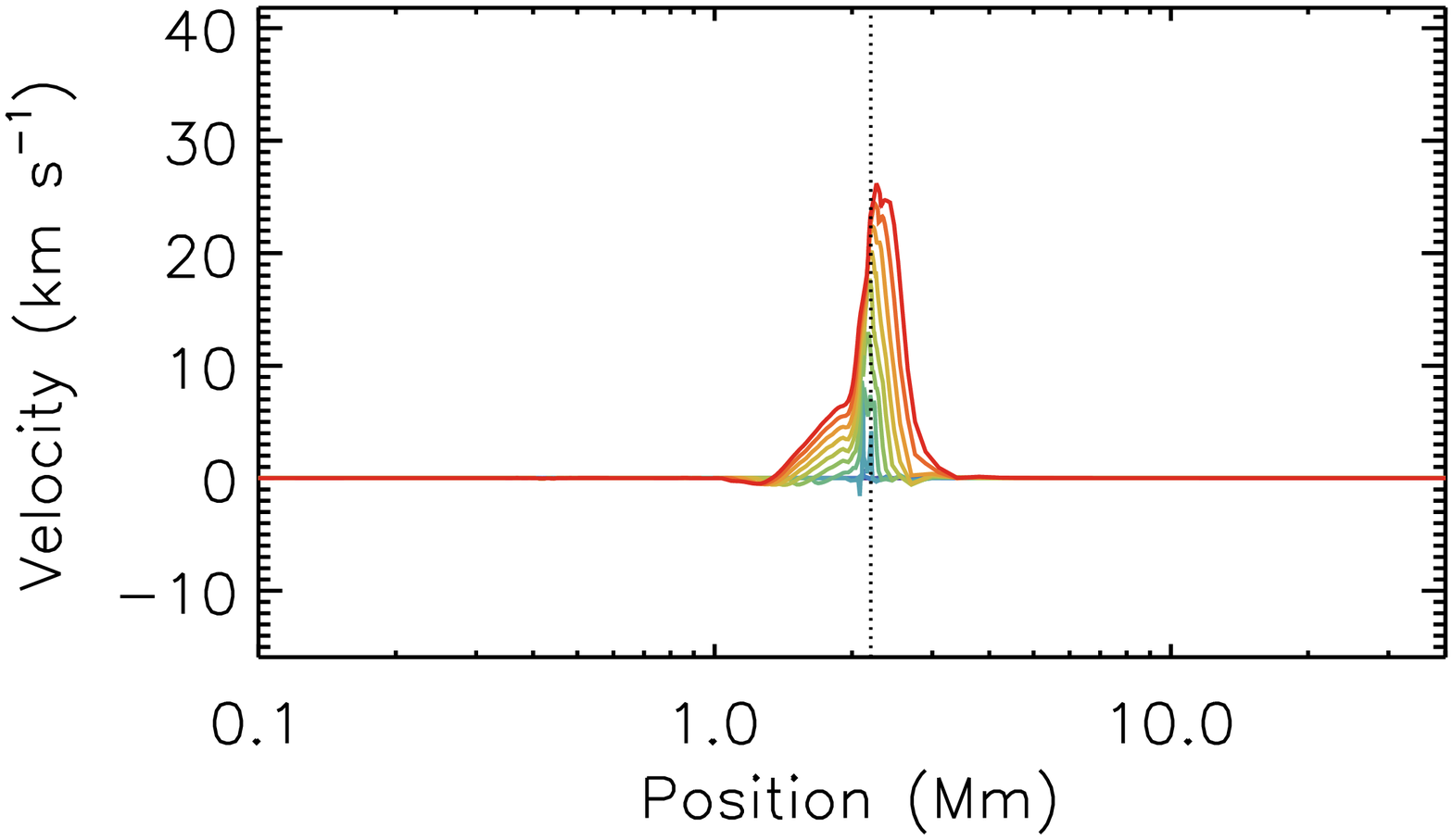}
\end{minipage}
\begin{minipage}[b]{0.5\linewidth}
\centering
\includegraphics[width=\textwidth]{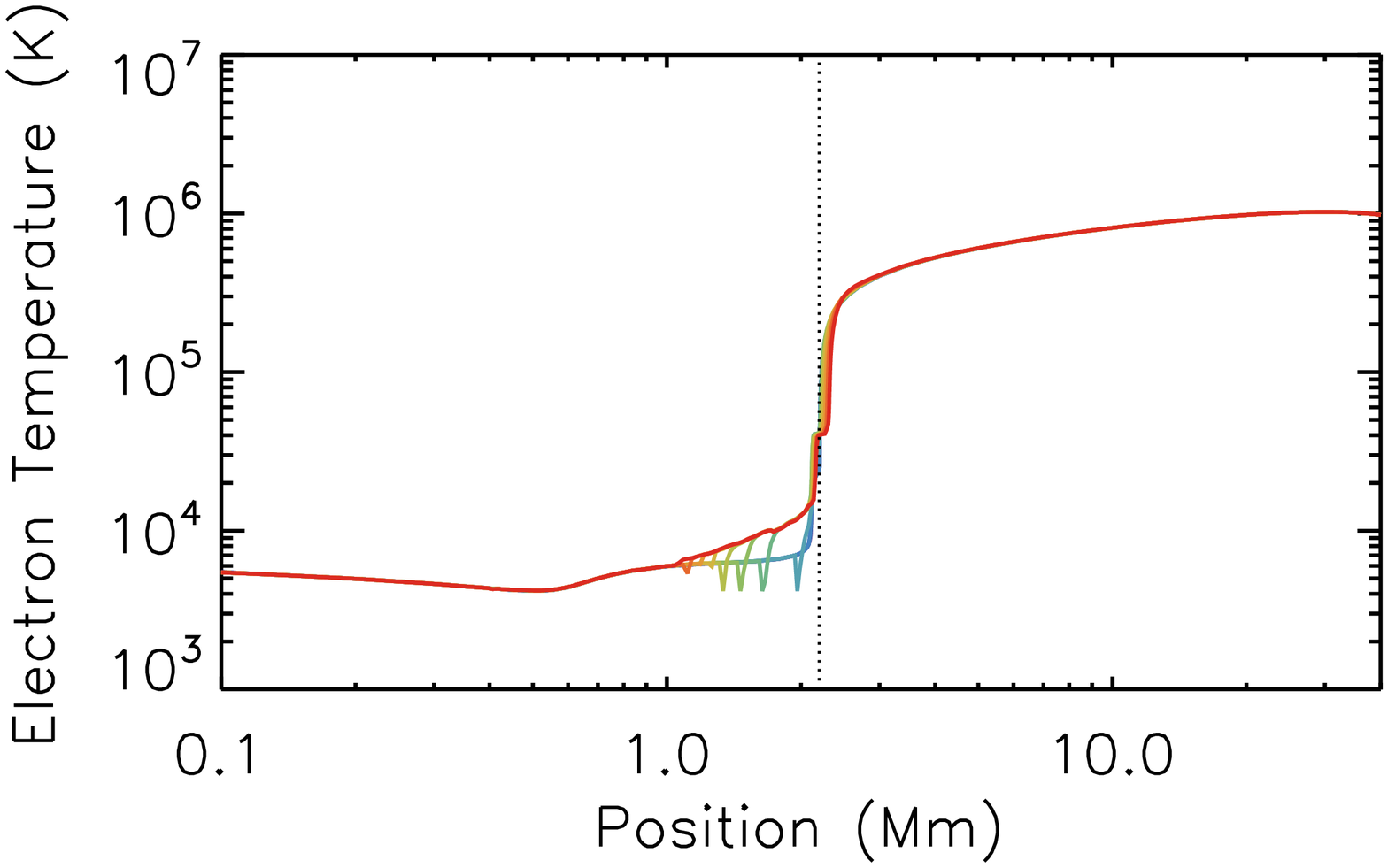}
\end{minipage}
\begin{minipage}[b]{0.5\linewidth}
\centering
\includegraphics[width=\textwidth]{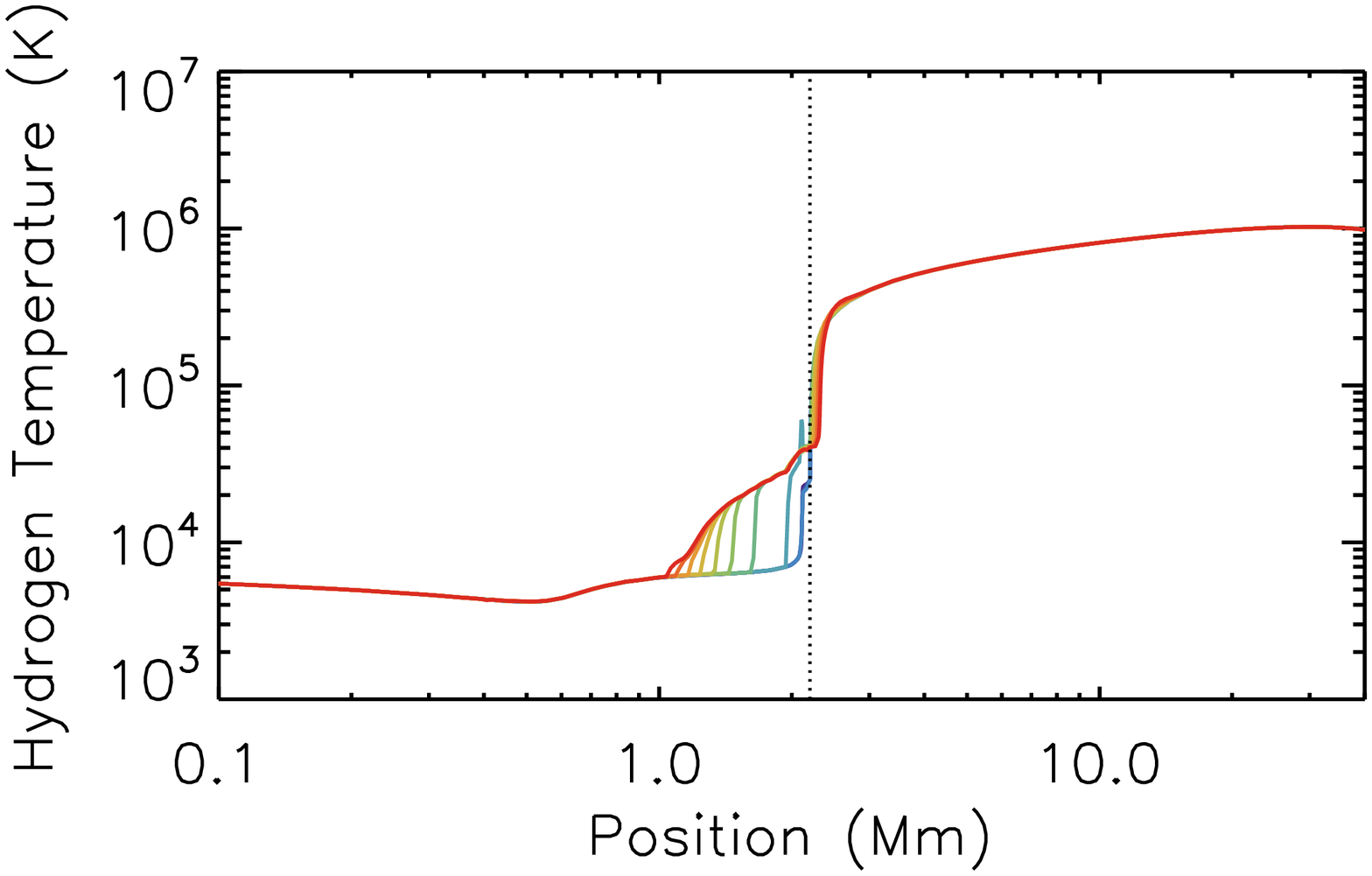}
\end{minipage}
\begin{minipage}[b]{0.5\linewidth}
\centering
\includegraphics[width=\textwidth]{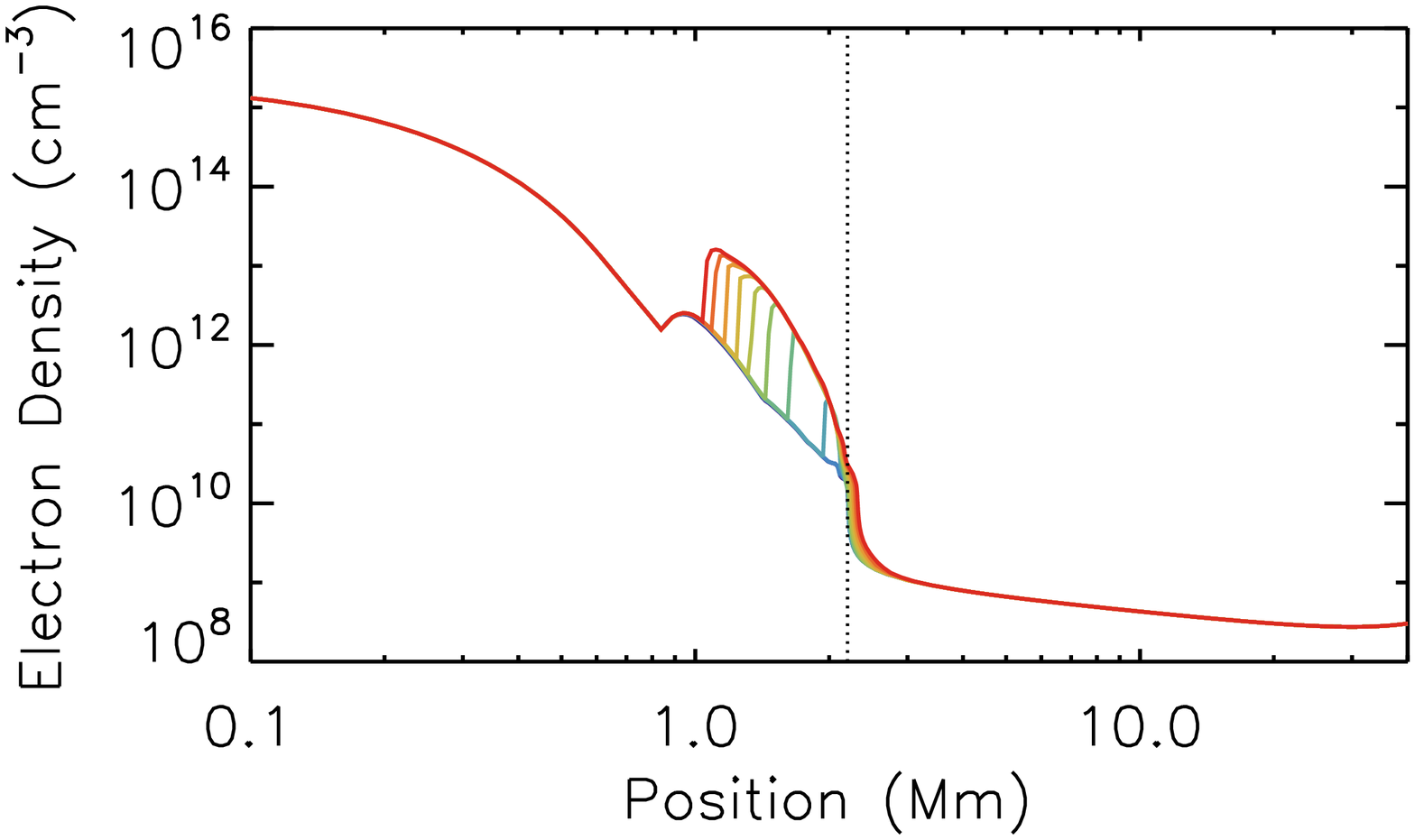}
\end{minipage}
\begin{minipage}[b]{0.5\linewidth}
\centering
\includegraphics[width=\textwidth]{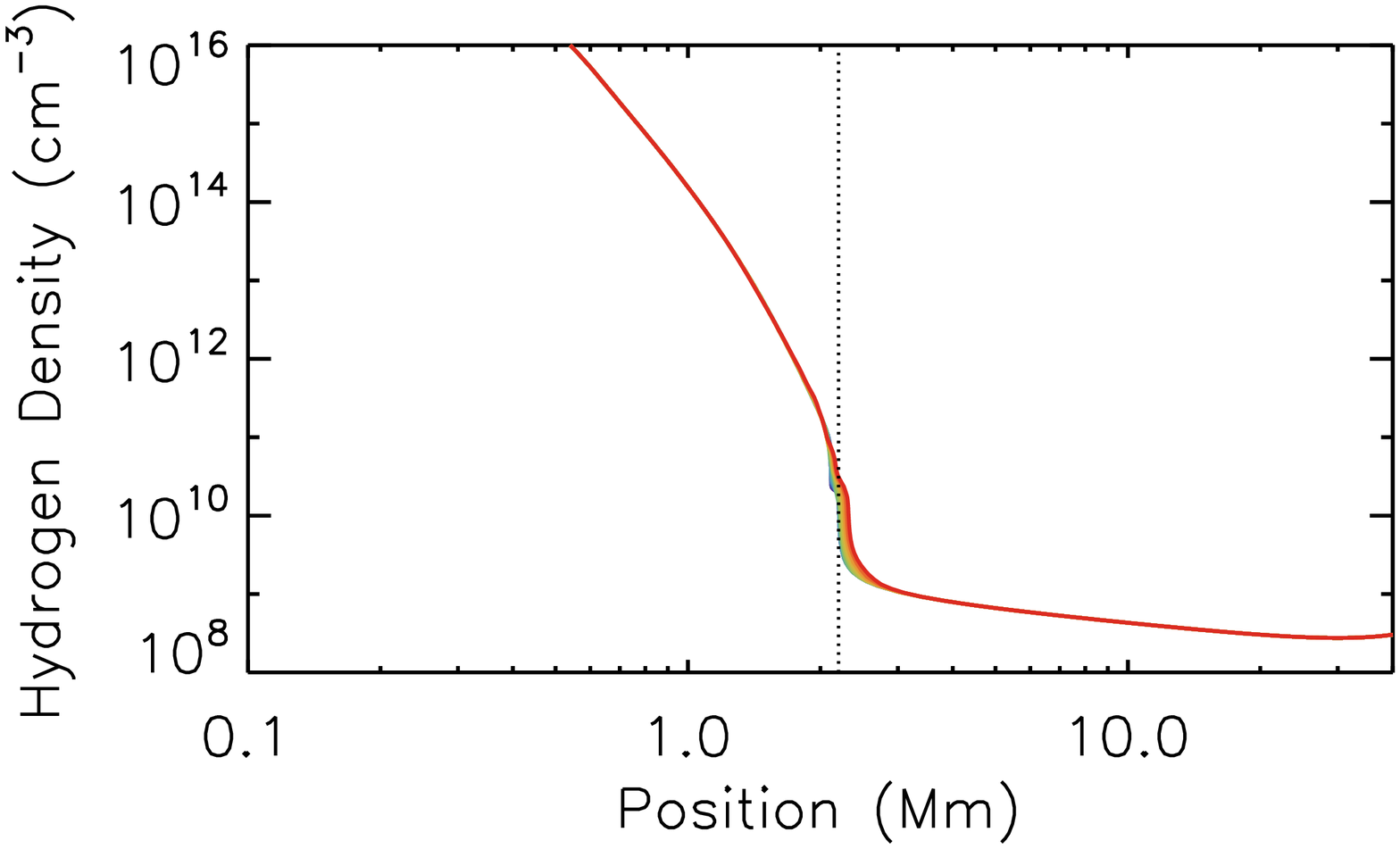}
\end{minipage}
\caption{Hydrodynamic variables for a pulse with duration 10 seconds, with $S(z_{0}) = 10^{10}$\,erg\,s$^{-1}$\,cm$^{-2}$, $k_{x,a} = 10^{-5}$\,cm$^{-1}$, and $f = 10$\,Hz (compare top row of Figure 1 in \citealt{reep2016}).  The different colors, going from violet to red, show the values at a one second cadence for the first ten seconds of the simulation.  Right-flowing velocities are defined as positive, left negative.  The black dashes on the heating plot marks the initial background heating level.  The plots have been truncated at a position just beyond the apex of the loop.}
\label{fig:RR1}
\end{figure*}

We now compare the new method to \citet{reep2016} in order to better understand the effect of propagation delays.  Figure \ref{fig:RR1} shows the hydrodynamic variables in a simulation with initial Poynting flux $S(z_{0}) = 10^{10}$\,erg\,s$^{-1}$\,cm$^{-2}$, perpendicular wave number at the apex $k_{x,a} = 10^{-5}$\,cm$^{-1}$, and frequency $f = 10$\,Hz, which can be compared to the top row of Figure 1 in \citet{reep2016}.  The colors show snapshots of the simulation every second, ranging from violet to red.  The x-axis is shown on a logarithmic scale to emphasize the chromosphere of the loop, which has a total length $2L = 60$\,Mm, so that the apex is at a position $30$\,Mm.  The first plot shows the total (electron + hydrogen) heating deposition (erg\,s$^{-1}$\,cm$^{-2}$) versus curvilinear position (with the background heating level included - the black dashed line), followed by the electron temperature (K), hydrogen temperature (K), bulk flow velocity (km\,s$^{-1}$), electron density (cm$^{-3}$), and total hydrogen density (cm$^{-3}$).  

\begin{figure*}
\begin{minipage}[b]{0.5\linewidth}
\centering
\includegraphics[width=\textwidth]{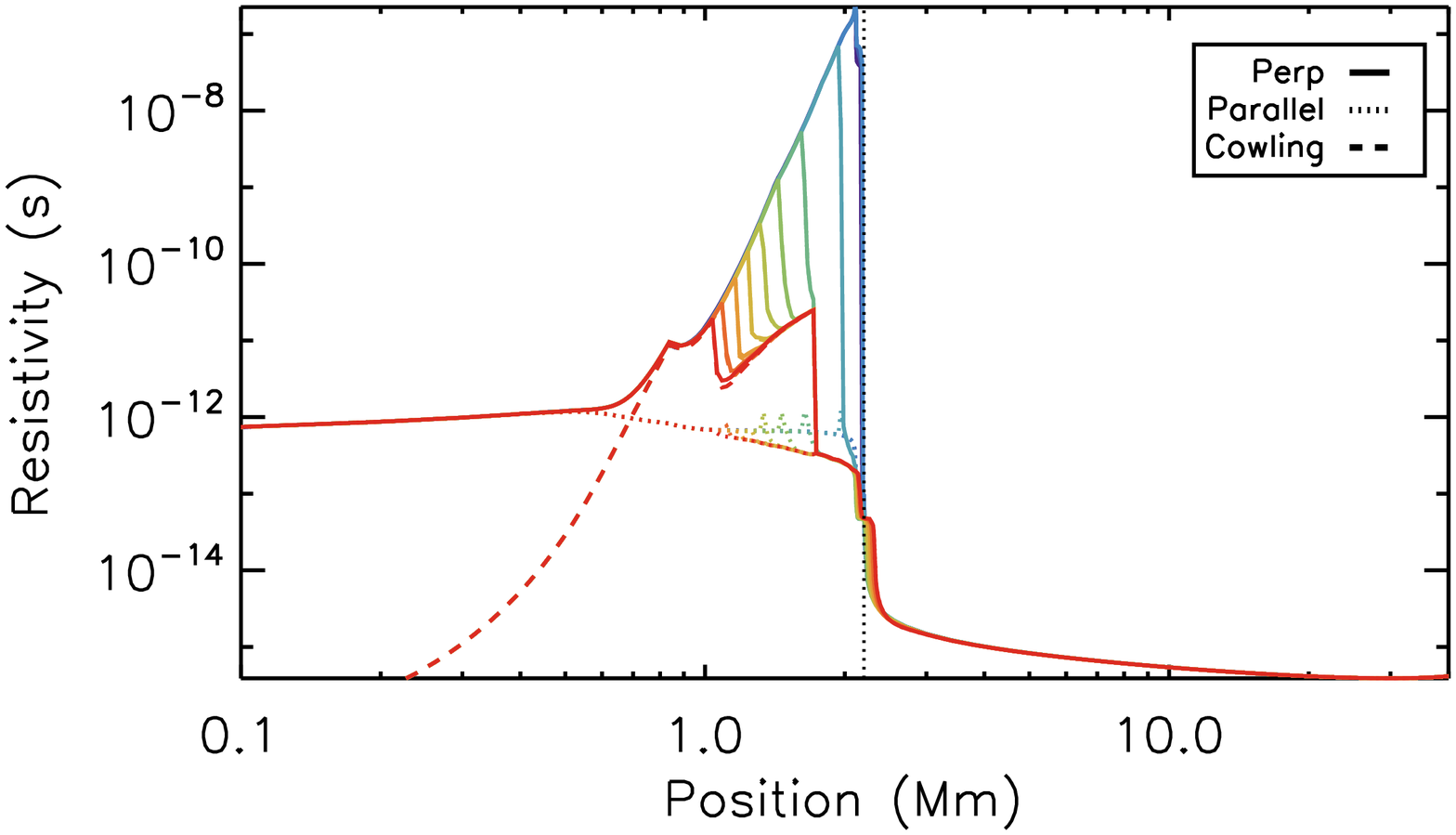}
\end{minipage}
\begin{minipage}[b]{0.5\linewidth}
\centering
\includegraphics[width=\textwidth]{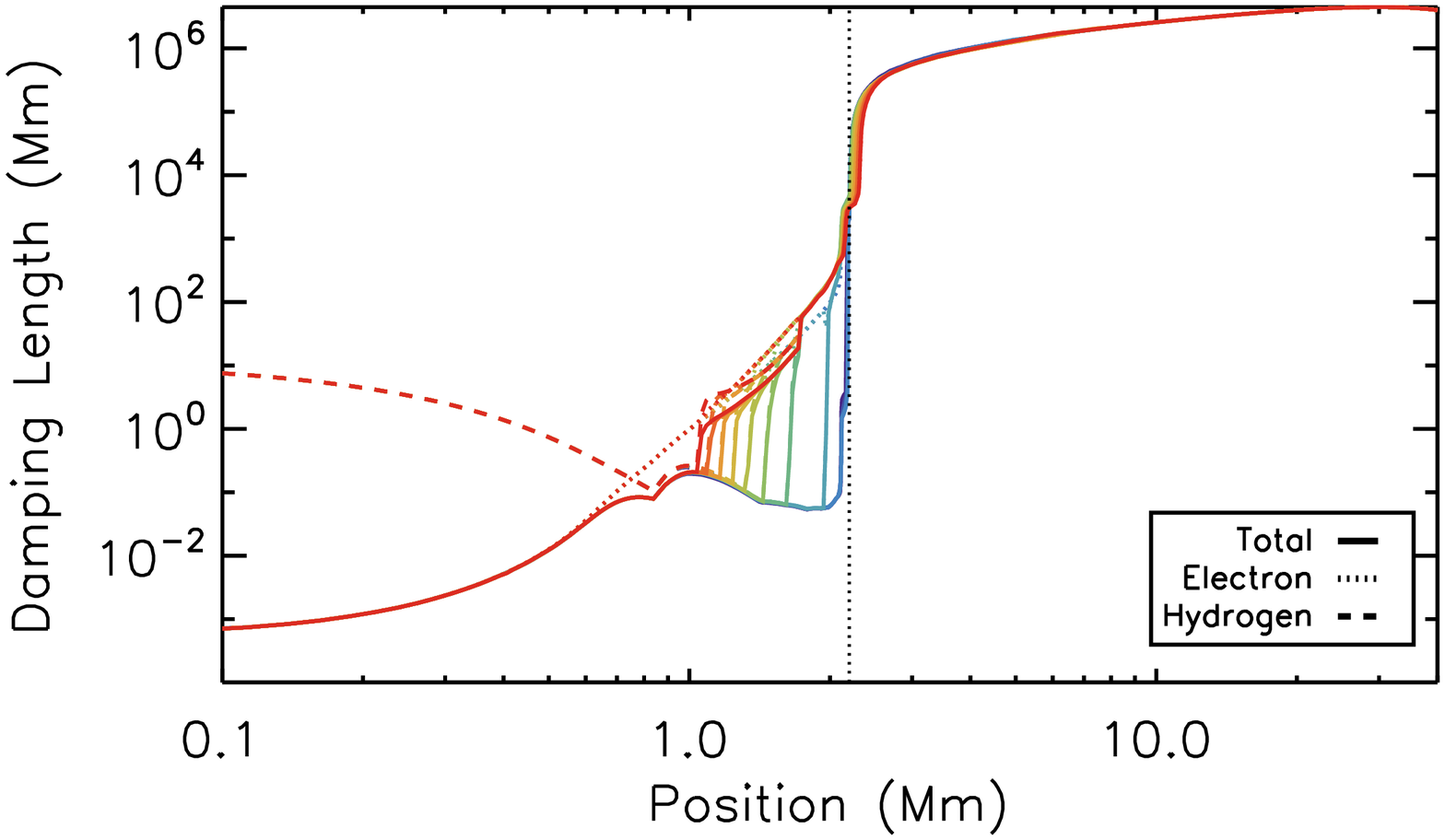}
\end{minipage}
\caption{The resistivities (left) and damping lengths (right) at the same times in the same simulation as Figure \ref{fig:RR1}.  As the pulse propagates downwards ionizing the plasma, the Cowling resistivity is reduced, thus causing damping due to electron collisions to become dominant at later times at the same depths.  As before, the colors show different times, at a 1 second cadence, from violet to red.}
\label{fig:RR1_damping}
\end{figure*}

It takes roughly 3 seconds for the wave to reach the chromosphere (with the assumed field strength), so that there is only weak coronal heating before that time.  After it impinges on the chromosphere, the pulse begins to dissipate, and the pulse slows as the Alfv\'en speed decreases due to an increasing density.  It takes roughly 7 seconds to travel 1\,Mm below the transition region, while the heating slowly grows with depth.  Since the perpendicular wave number is small, the Cowling resistivity initially dominates the energy dissipation terms, so that the hydrogen temperature rises more than the electron temperature, though both only increase modestly.  As the pulse propagates, the plasma is ionized, reducing the Cowling resistivity, so that at later times damping due to electron collisions becomes the dominant term.  Figure \ref{fig:RR1_damping} shows the resistivities and damping lengths in the simulation, demonstrating the reduction in Cowling resistivity.  This indicates the importance of the ionization level in determining the depth and rate at which waves dissipate.  The parallel resistivity also decreases slightly as the electron density increases due to the change in ionization.  As with the simulation in \citet{reep2016}, the bulk flow velocity evaporates gently.  As the chromosphere heats, the ionization fraction rises, locally raising the electron density, while the weak evaporation carries a small number of both electrons and ions into the corona. 

\begin{figure*}
\begin{minipage}[b]{0.5\linewidth}
\centering
\includegraphics[width=\textwidth]{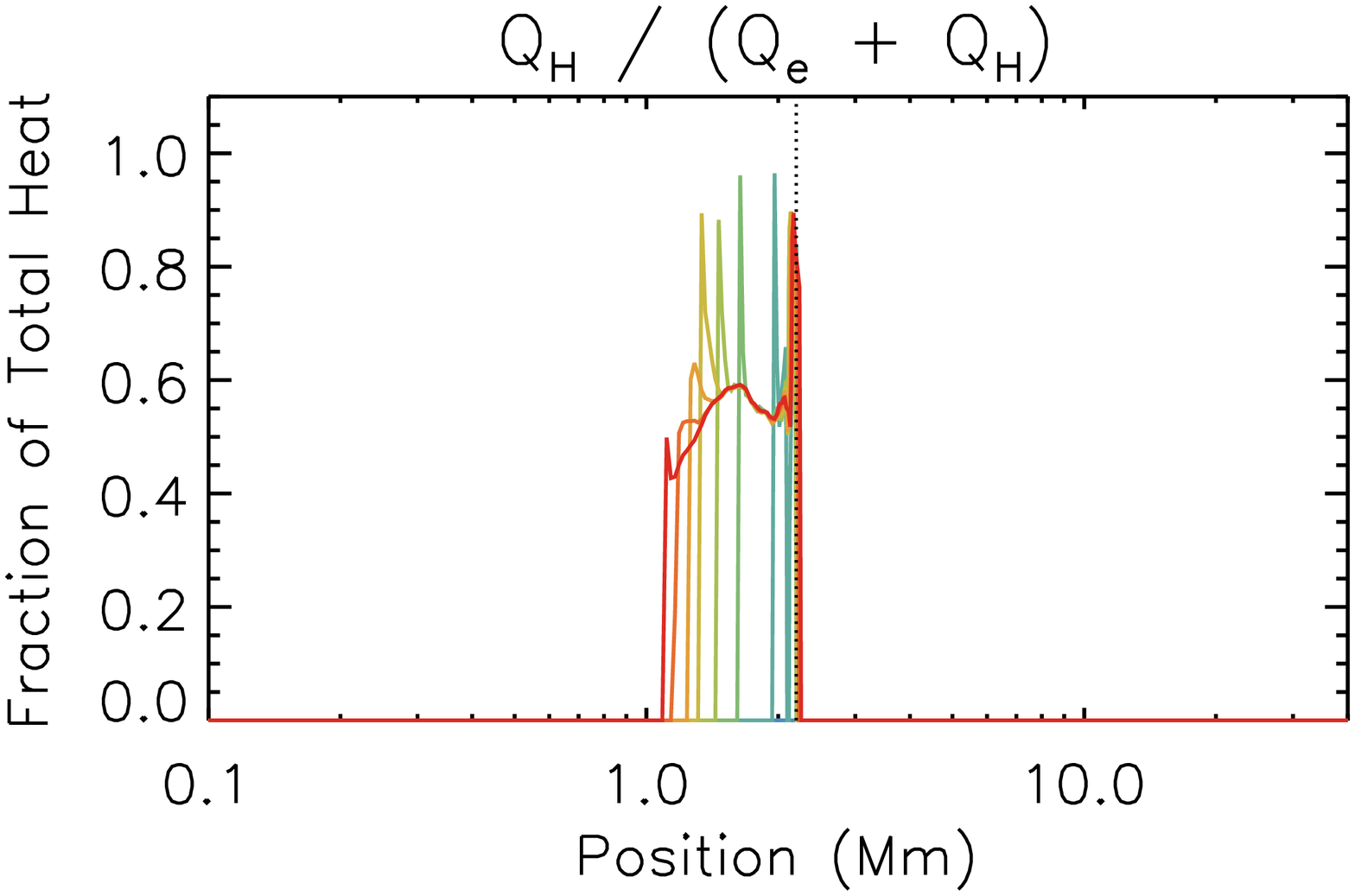}
\end{minipage}
\begin{minipage}[b]{0.5\linewidth}
\centering
\includegraphics[width=\textwidth]{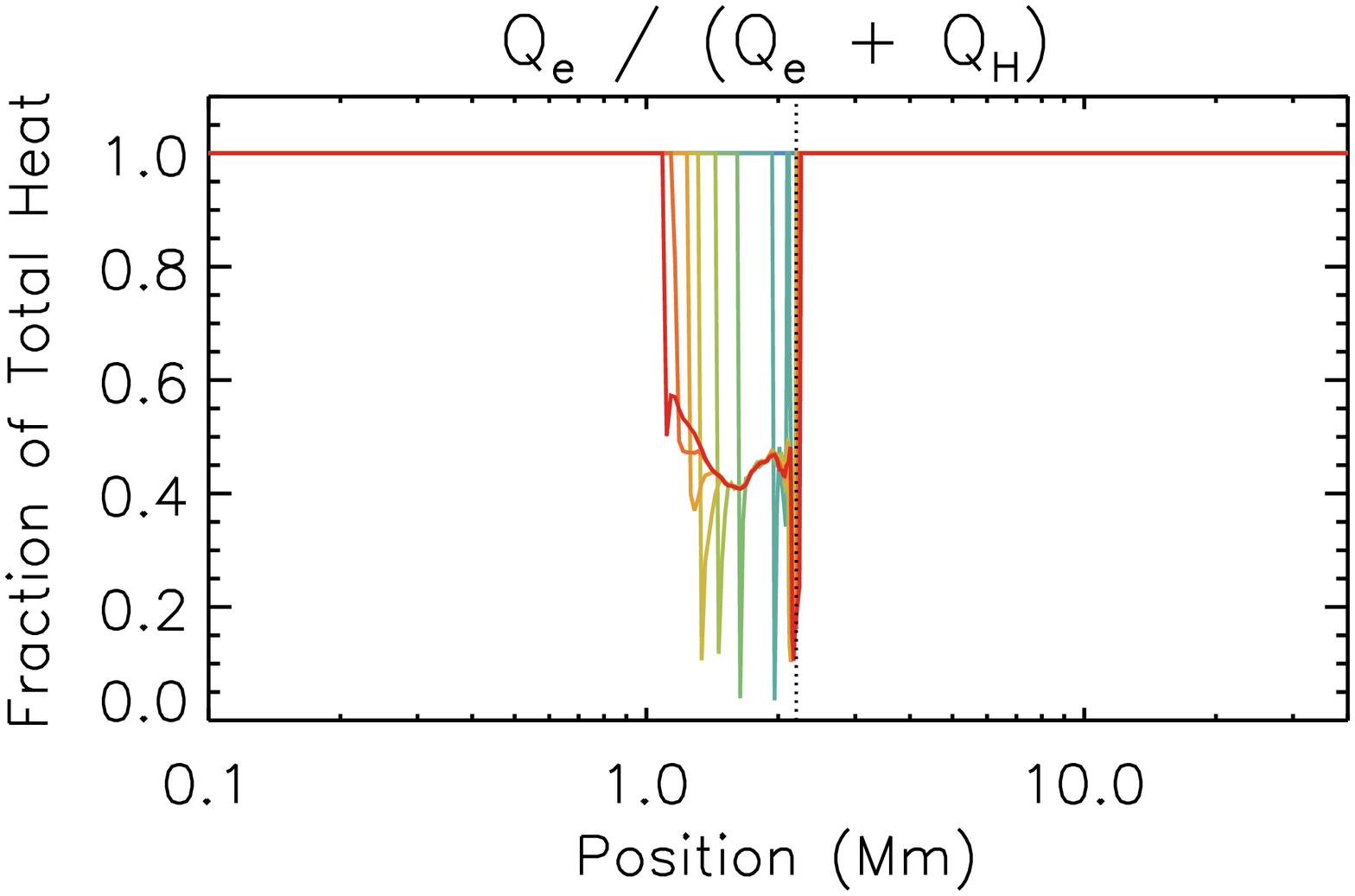}
\end{minipage}
\caption{The partition of the heat as a function of position and time for the simulation in Figure \ref{fig:RR1}.  The left plot shows the fraction of dissipated energy that heats hydrogen, the right plots shows the fraction that heats electrons.  The background heating is assumed to go entirely into electrons.}
\label{fig:partition}
\end{figure*}

For this same simulation, we show the partition of heat into hydrogen and electrons in Figure \ref{fig:partition}.  The left plot shows the fraction of dissipated energy that heats the hydrogen, while the right plot shows the fraction that heats the electrons.  The background heating is assumed to go entirely into electrons.  In this case, since the perpendicular wave number $k_{x}$ is small, a significant fraction of the energy goes into heating the hydrogen directly, reaching above 80\% at the leading edge of the pulse, and remaining close to 50\% trailing that.    

\begin{figure*}
\begin{minipage}[b]{0.5\linewidth}
\centering
\includegraphics[width=\textwidth]{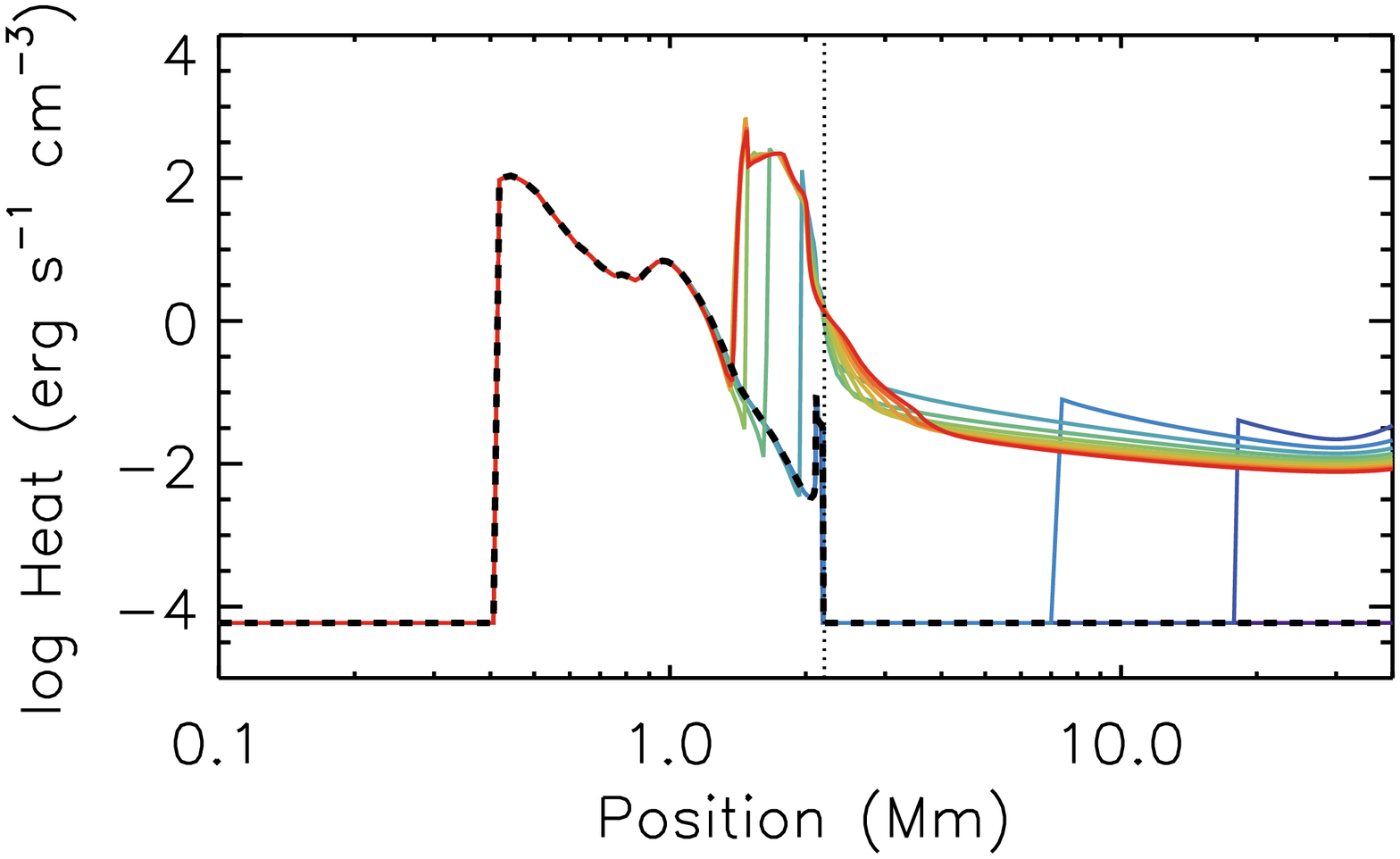}
\end{minipage}
\begin{minipage}[b]{0.5\linewidth}
\centering
\includegraphics[width=\textwidth]{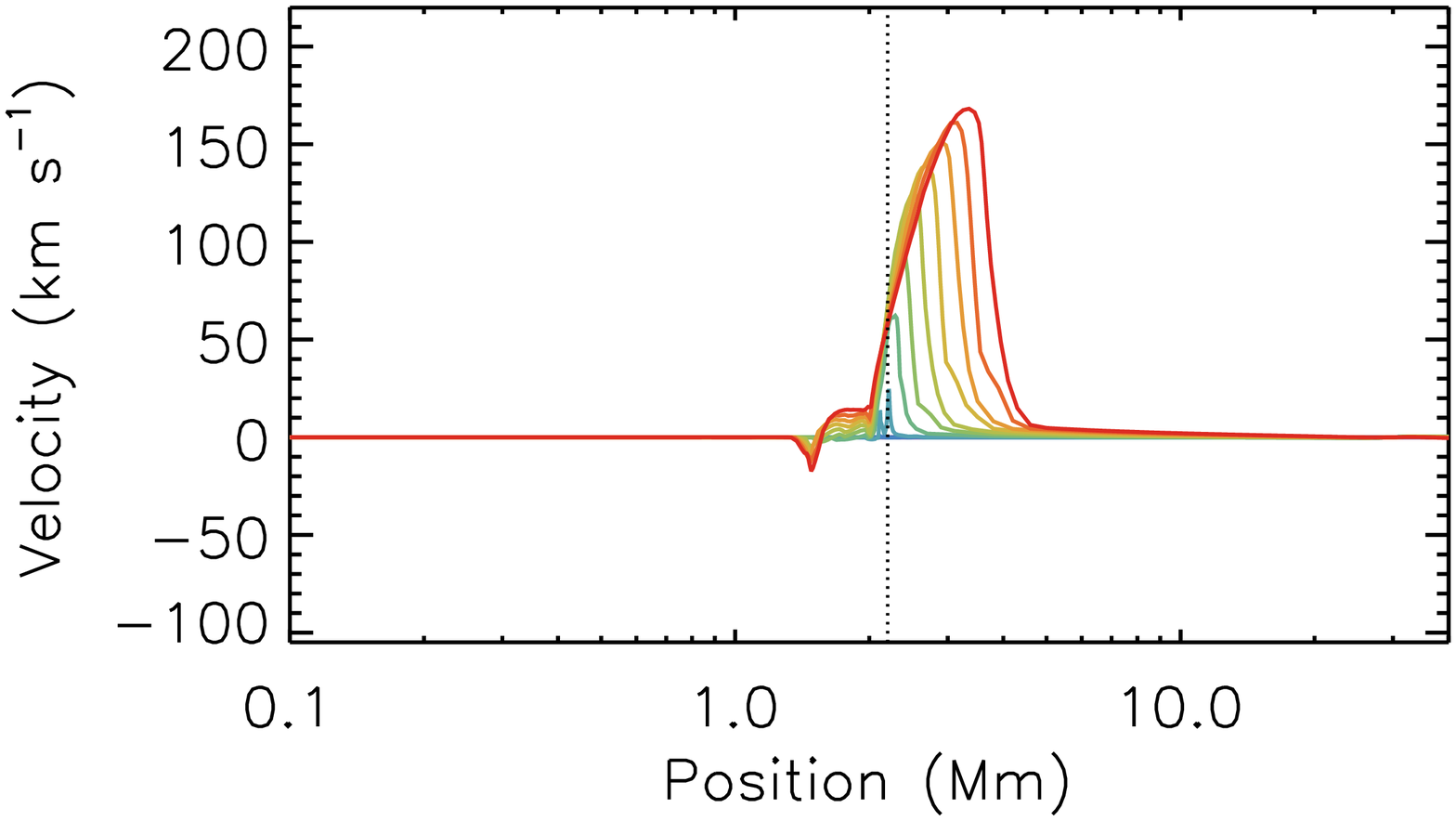}
\end{minipage}
\begin{minipage}[b]{0.5\linewidth}
\centering
\includegraphics[width=\textwidth]{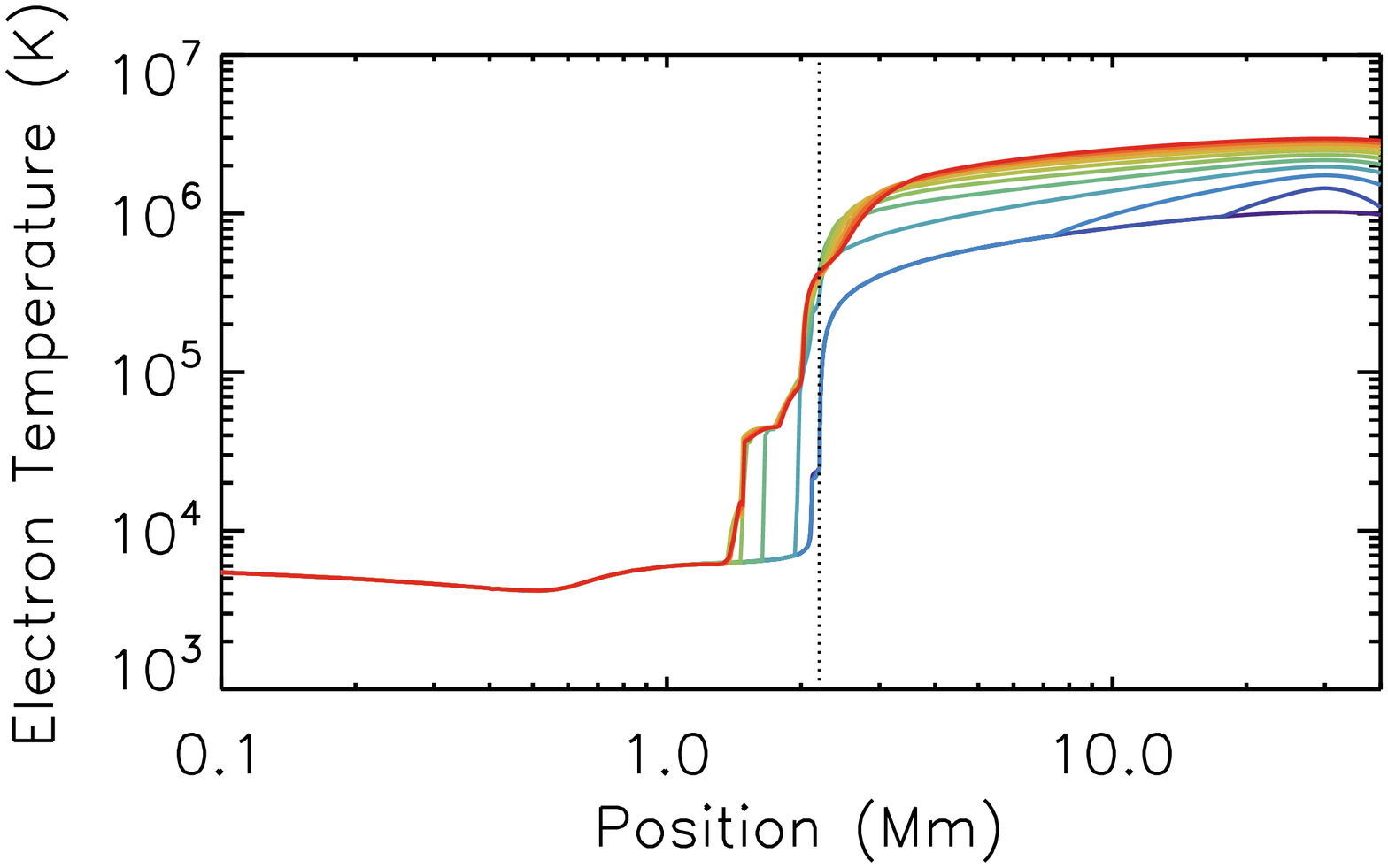}
\end{minipage}
\begin{minipage}[b]{0.5\linewidth}
\centering
\includegraphics[width=\textwidth]{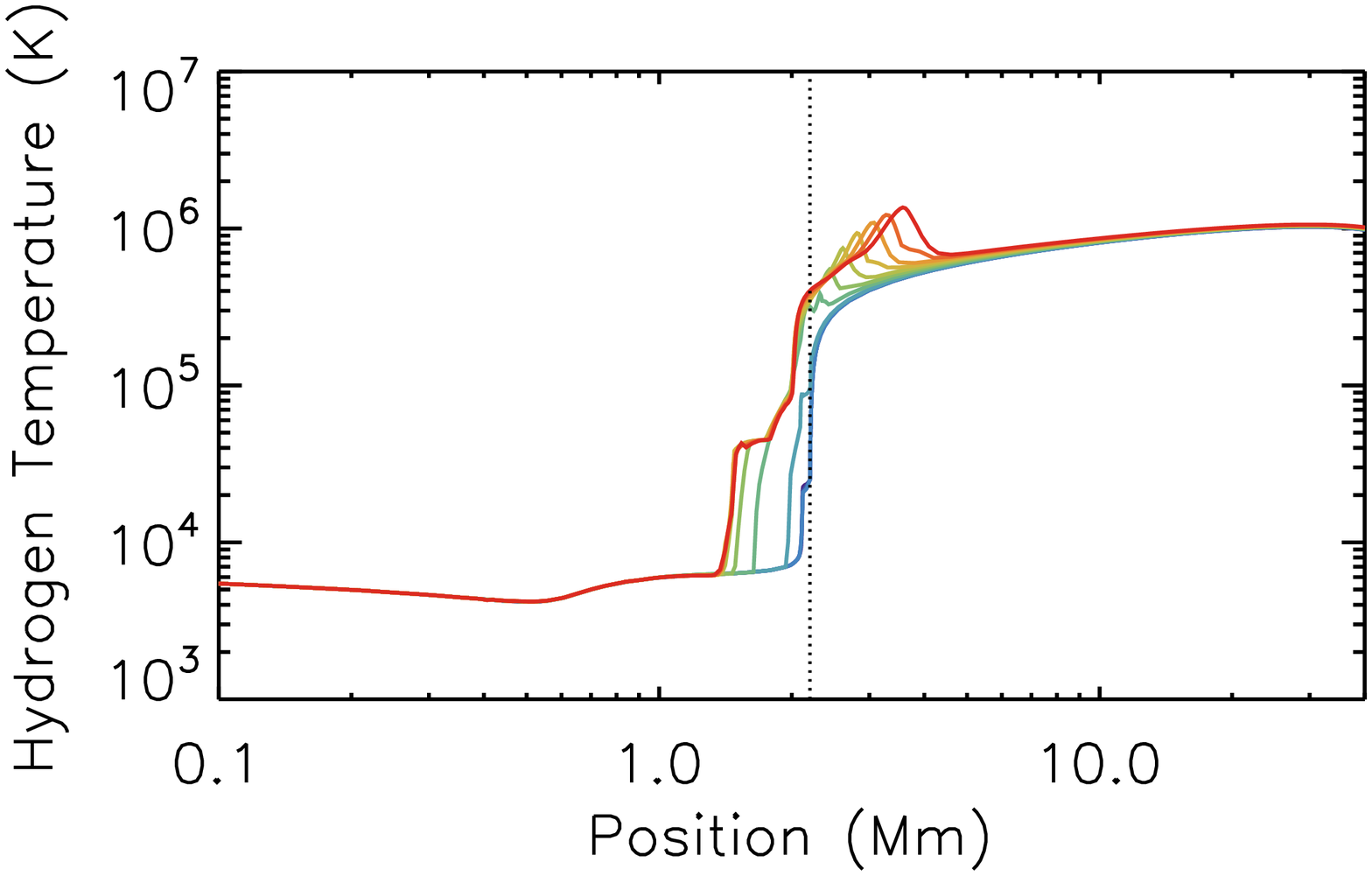}
\end{minipage}
\begin{minipage}[b]{0.5\linewidth}
\centering
\includegraphics[width=\textwidth]{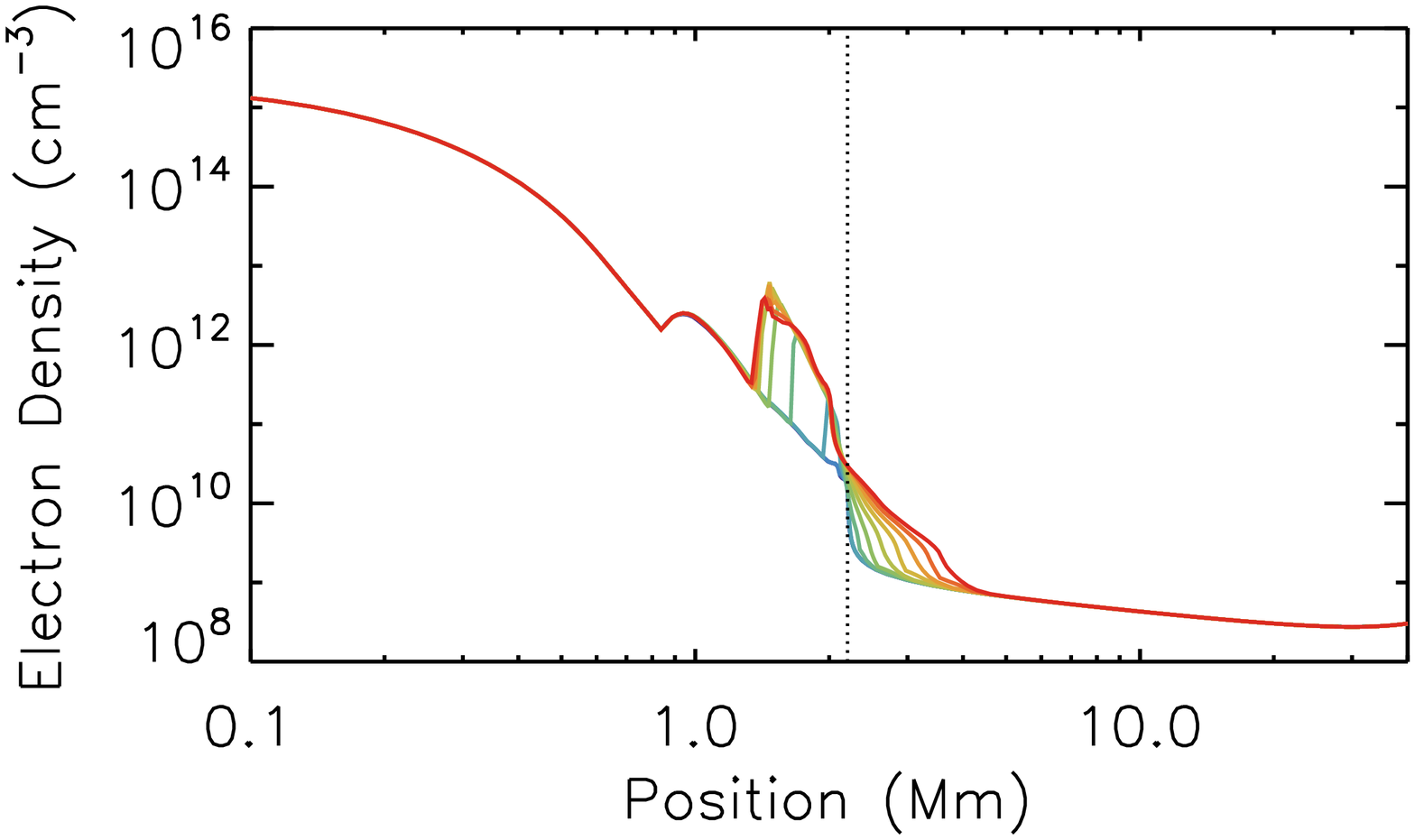}
\end{minipage}
\begin{minipage}[b]{0.5\linewidth}
\centering
\includegraphics[width=\textwidth]{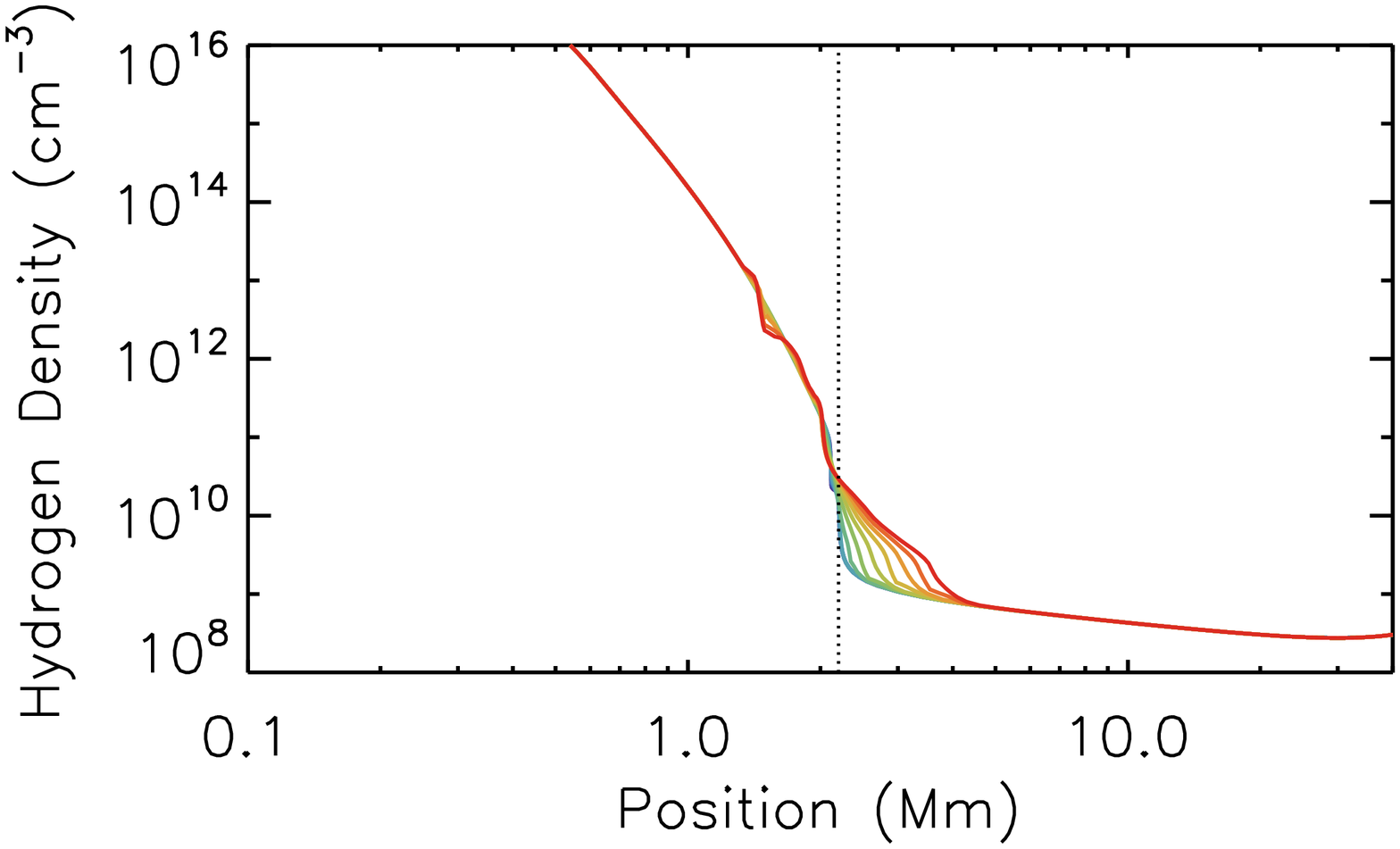}
\end{minipage}
\caption{Similar to Figure \ref{fig:RR1}, with $k_{x,a} = 4 \times 10^{-4}$\,cm$^{-1}$ and $f = 10$\,Hz (compare middle row of Figure 1 in \citealt{reep2016}).}
\label{fig:RR2}
\end{figure*}

In Figure \ref{fig:RR2}, we show a simulation with a higher perpendicular wave number $k_{x,a} = 4 \times 10^{-4}$\,cm$^{-1}$ (but otherwise equal parameters), which increases the relative importance of parallel dissipation.  Due to the rise in parallel dissipation (both ion-electron and electron-neutral collisions), the heating is stronger in both the corona and upper chromosphere, and peaks more sharply.  As a result, the electron temperature rises all across the corona, spreading outwards from the apex of the loop at $z = 30$\,Mm and into the chromosphere.  The hydrogen temperature rises sharply here as well, driving strong, explosive flows, and comparable to the same case in \citet{reep2016} (middle row of Figure 1 in that paper).  Since the energy is deposited higher in the chromosphere, the waves do not travel as deep before dissipating, so the rise in ionization does not occur as deep as in the previous case.

As mentioned before and shown in Figure \ref{fig:RR1_damping}, as a pulse propagates into the chromosphere and dissipates its energy, the ionization fraction and electron density grow, thus decreasing both the Cowling and parallel resistivities.  The trailing end of a long duration pulse, or any subsequent pulses, therefore propagate to greater depths into the chromosphere than the leading edge (or pulse) because of that reduction in the resistivity.  Effectively, earlier pulses bore a hole into the chromosphere that allows later ones to penetrate deeper and deeper, even at frequencies expected to strongly dissipate high in the initial chromosphere.  Heating due to the dissipation therefore effectively propagates downwards.  

This behavior is {\it opposite} to an electron beam.  As a beam deposits its energy in the chromosphere, driving evaporation into the corona, the mean free path of any later electrons is significantly reduced due to the rise in coronal density.  The result is that the location of energy deposition for long duration beams rises into the corona and becomes more and more localized near the injection site, as shown for example in Figure 3 of \citet{reep2015}.  

There is therefore an important distinction in the heating profiles between the case of Alfv\'en waves and electrons beams.  Electrons propagating down a loop are stopped as they collide with ambient plasma, in particular as the density rises in the chromosphere, thus shortening their mean-free paths.  Before there is significant evaporation into the corona, electrons deposit their energy at an approximately constant depth, as their mean-free paths do not change (assuming there are no significant time-varying changes in the beam, such as an increase in the low energy cut-off).  This is in stark contrast to the case of Alfv\'en waves, which propagate significantly more slowly than beams, and deposit energy across more of the chromosphere as the density rises.  The difference in energy deposition locations causes differences in the pressure, ionization, and flows.

\begin{figure*}
\begin{minipage}[b]{\linewidth}
\centering
\includegraphics[width=\textwidth]{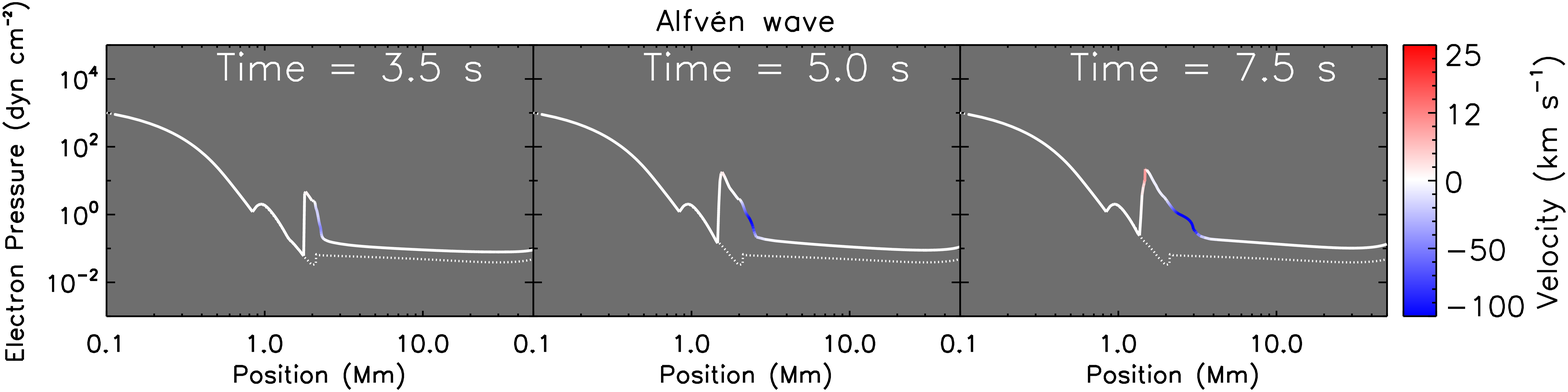}
\end{minipage}
\begin{minipage}[b]{\linewidth}
\centering
\includegraphics[width=\textwidth]{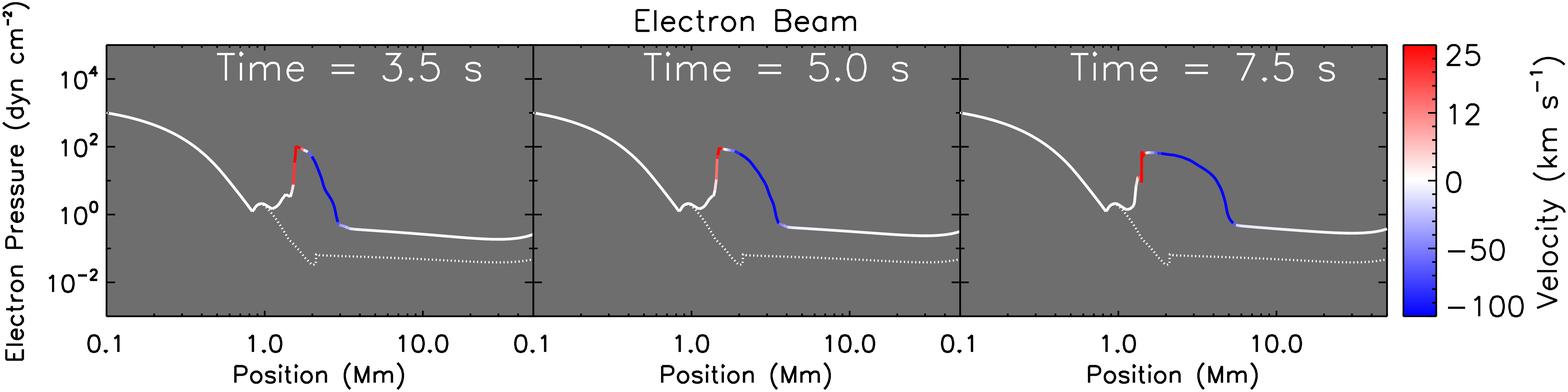}
\end{minipage}
\caption{A comparison of the development of the pressure profiles of loops heated by an Alfv\'en wave (top) and electron beam (bottom).  In the case of the wave, because it propagates increasingly deep, the spike in pressure propagates downwards.  In the case of the beam, the electrons reach essentially the same depth continually, so that the pressure spike does not move.  The colors on this plot refer to the local velocity, colored blue where the plasma is up-flowing, red where it is down-flowing, and white where it is stationary.  Down-flowing velocities are defined as positive in this and the next figure.  The white dotted line shows the initial profile.}
\label{fig:beamcomp_pressure}
\end{figure*}

We can directly compare the behavior for beam and wave heating.  In Figure \ref{fig:beamcomp_pressure}, we show the electron pressure profiles as a function of position at three select times for an Alfv\'en wave simulation with $f = 10$\,Hz, $k_{x} = 4 \times 10^{-4}$\,cm$^{-1}$, and $S(z_{0}) = 10^{10}$\,erg\,s$^{-1}$\,cm$^{-2}$ (top row) and an electron beam with low energy cut-off $E_{c} = 10$\,keV, spectral index $\delta = 5$, and energy flux $F_{0} = 10^{10}$\,erg\,s$^{-1}$\,cm$^{-2}$ (bottom row), using the heating form derived by \citet{emslie1978}.  We have colored the locations by the local flows, blue where the plasma is up-flowing, red where it is down-flowing, and white where it is stationary.  The white dots indicate the initial profile.  The spike in the chromospheric pressure propagates downwards in the case of the waves, whereas in the case of the beam it does not reach progressively greater depths.  In both cases, a pressure front begins to rise into the corona and evaporate plasma, although it travels at a higher speed for the beam.  

\begin{figure*}
\begin{minipage}[b]{\linewidth}
\centering
\includegraphics[width=\textwidth]{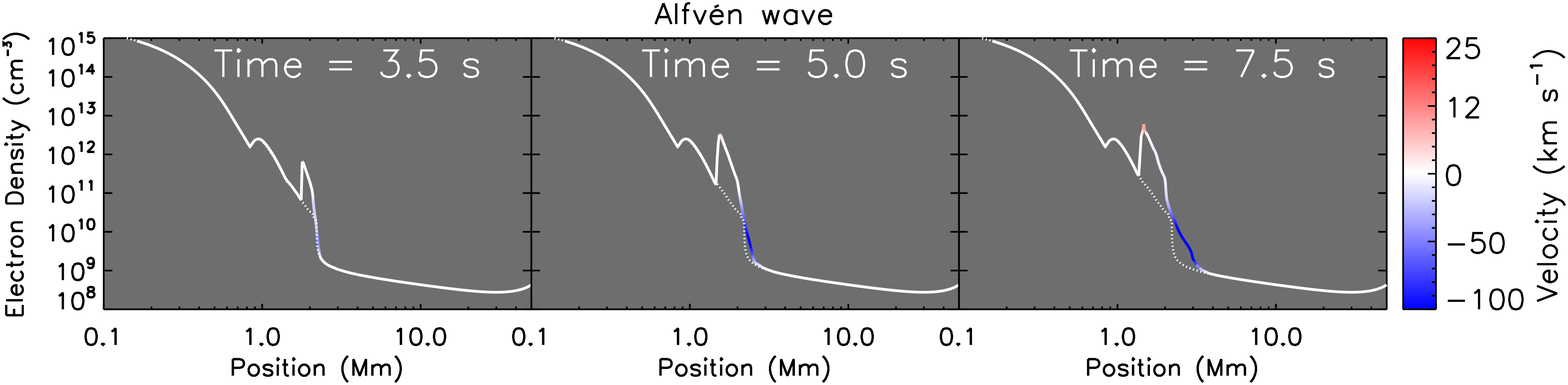}
\end{minipage}
\begin{minipage}[b]{\linewidth}
\centering
\includegraphics[width=\textwidth]{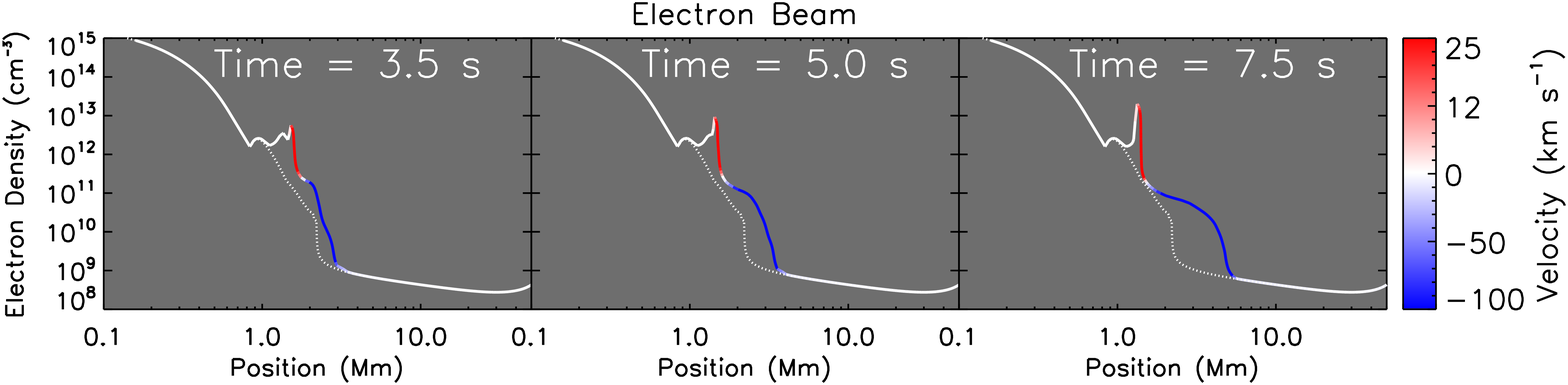}
\end{minipage}
\caption{The electron density profiles for the same simulations in Figure \ref{fig:beamcomp_pressure}, shown similarly.  As the waves propagate downwards, they ionize the plasma, increasing the electron density at greater and greater depths, whereas a beam reaches essentially the same depth at all times.  }
\label{fig:beamcomp_density}
\end{figure*}

In Figure \ref{fig:beamcomp_density}, we show the electron density profiles in a similar manner to the previous figure.  As noted before, as Alfv\'en waves propagate to deeper depths, they cause a downward propagating ionization spike, thus raising the local electron density while simultaneously driving up-flows into the corona.  In the case of a beam, however, the plasma at the depth of energy deposition becomes ionized, but there is no propagation of that ionization spike.  

These differences in the ionization, flows, and pressures suggest stark differences in chromospheric line profiles.  Using the instantaneous method, \citet{kerr2016} found that the \ion{Mg}{2} k line behaved differently in an electron beam simulation from an Alfv\'en wave simulation.  Specifically, the central reversal of the line gradually disappears, and the line develops a strong red-blue asymmetry in the wave simulation, while the line remains roughly symmetric with a strong central reversal during the electron beam simulation.  On the other hand, the \ion{Ca}{2} 8542\,\AA\ line was similar in both simulations, with a small difference in the Doppler shifts.  The inclusion of travel time effects would likely exacerbate the differences found by those authors, which warrants a deeper investigation.  Other authors have similarly reported discrepancies in the modeling of the \ion{Mg}{2} lines when modeled with an electron beam (\textit{e.g.} \citealt{rubiodacosta2016}), which warrants further examination.  Because these properties all vary with chromospheric depth and with time, in principle they could be probed by high cadence spectrometry ($\lesssim 1$\,s) to track the propagation of these waves and to measure the hydrodynamic profiles in time.  We plan to address these possibilities in future work.  

\section{Conclusions}  
\label{sec:conclusions}

In this work, we have updated the model from \citet{reep2016} to include travel time effects due to the propagation of Alfv\'enic waves.  The method is sufficiently general to be implemented in any hydrodynamics model that does not solve the MHD wave equations.  To test that this method is valid, we have compared it against the MHD code Lare3D, adapted for a field-aligned flux tube, and found a generally good agreement.  

We have drawn a number of important conclusions from this work:

\begin{enumerate}
\item[(1)] Alfv\'enic waves can heat all levels of the chromosphere, and the damping of these waves depends strongly on not only the wave parameters, but also the ionization level.  Ion-neutral friction is extremely efficient at damping waves, so a high proportion of neutral atoms greatly increases the damping of Alfv\'en waves, while a low proportion means that waves travel mostly unimpeded.
\item[(2)] Because waves ionize the plasma as they propagate through the chromosphere, early waves effectively bore a hole in the chromosphere through which later waves can more readily propagate without significant dissipation.  Successive waves penetrate to ever greater depths, and high frequency waves penetrate significantly deeper than expected from our previous paper.  
This contrasts directly with the well-known result for electron beams: as evaporation brings material into the corona, later electrons have a shortened mean-free path.  The height at which beams deposit their energy therefore tends to travel upwards along the loop for long duration heating (see {\it e.g.} \citealt{reep2015}).  
\item[(3)] The propagation of these waves causes pressure and ionization fronts to form in the chromosphere that propagate with time.  This is in contrast to an electron beam, where the depth of the maximum pressure remains localized.  To observe this difference would likely require high cadence observations due to the large Alfv\'en speed.  Since there are differences in the chromospheric condensations, there would likely also be differences in spectral line profiles (\textit{e.g.} red-blue asymmetry) that could be measured.  Further investigation of this is warranted.
\item[(4)] Alfv\'en wave heating has unique observational signatures.  These signatures should be used to start determining the relative contribution of waves and electron beams to flare heating, nanoflare heating, or more generally, the partitioning of the energy released by reconnection.
\end{enumerate}

We will continue to develop this method in future work.  Reflection, mode conversion, the ponderomotive force, and other effects require proper treatment in an accurate model, and these would benefit from an in-depth MHD study.  Further, we have predicted that the signatures of wave dissipation in the chromosphere could be detected with high cadence spectroscopy, so both observational and modeling studies of chromospheric lines should be undertaken.  A study including a full wave spectrum ({\it e.g.} \citealt{tarr2017}) might also provide insight about the heating processes occurring in the chromosphere.

\appendix
\label{appendix}

In order to demonstrate the current limitation of this method while neglecting reflection, we briefly examine one more comparison between HYDRAD and Lare.  We examine a simulation with one pulse with frequency $f = 1$\,Hz, where the reflection coefficient at the transition region is expected to be around 45--50\% (Figure \ref{fig:trans}).  We set the initial Poynting flux to $S_{0} = 10^{9}$\,erg\,s$^{-1}$\,cm$^{-2}$ and the perpendicular wave number $k_{x} = 0$ as before.  The initial conditions are the same as in Section \ref{sec:larecomparison}.

\begin{figure*}
\begin{minipage}[b]{0.5\linewidth}
\centering
\includegraphics[width=\textwidth]{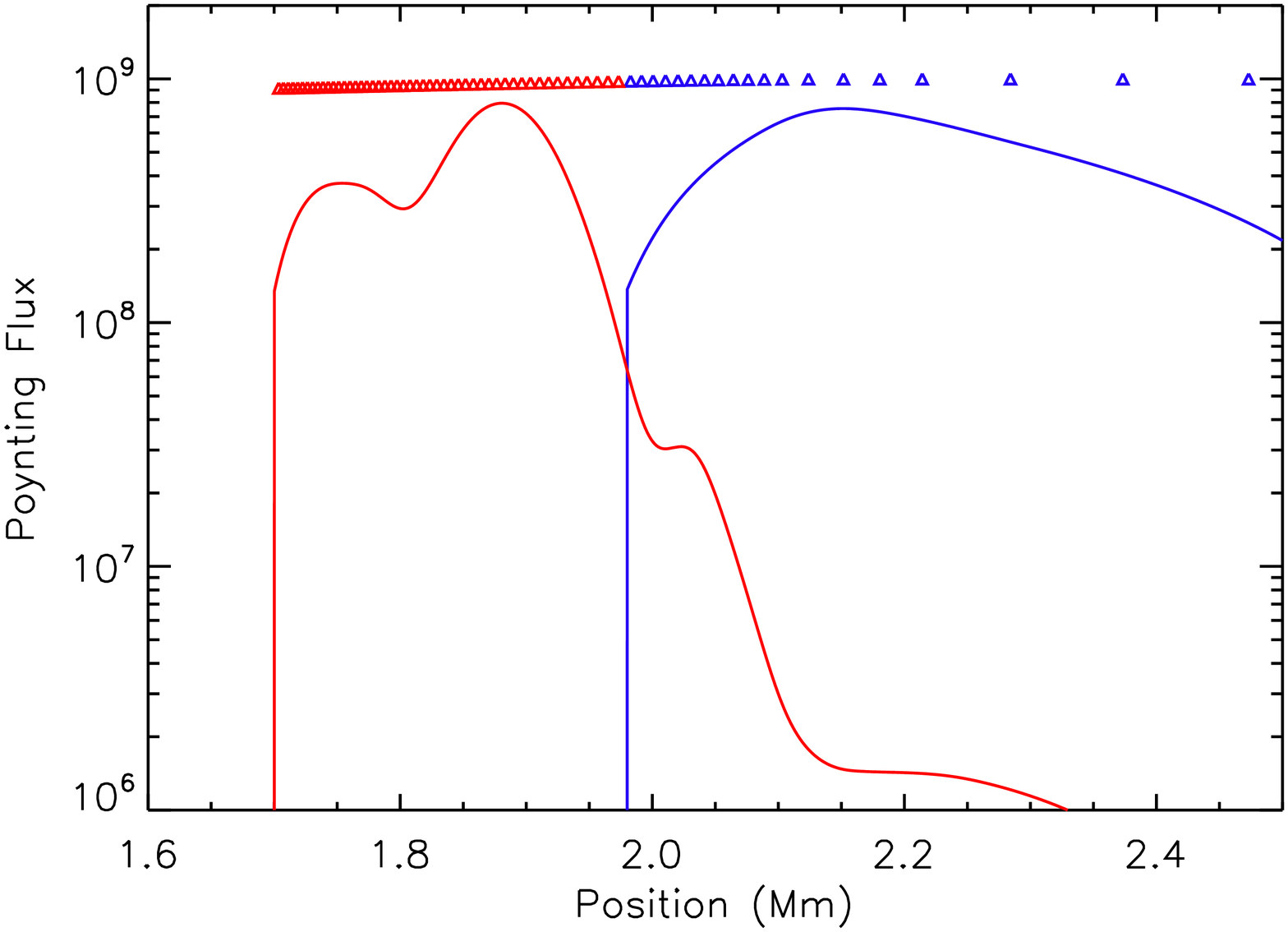}
\end{minipage}
\begin{minipage}[b]{0.5\linewidth}
\centering
\includegraphics[width=\textwidth]{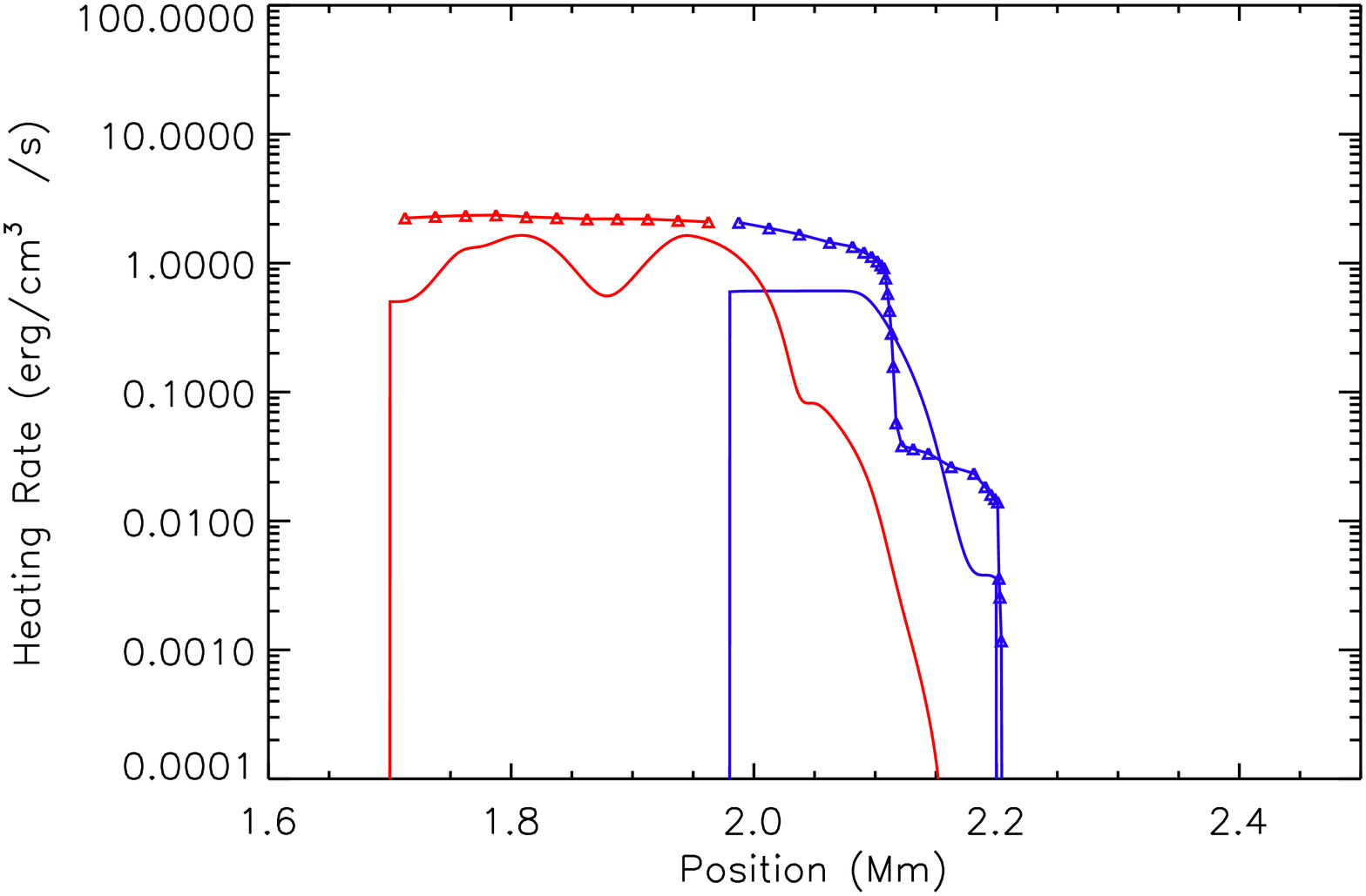}
\end{minipage}
\caption{A comparison between a wave simulation with HYDRAD (points) and Lare (lines) in the case of strong reflection, $f = 1$\,Hz and initial Poynting flux of $10^{9}$\,erg\,s$^{-1}$\,cm${-2}$.  Similar to Figure \ref{fig:LC_sim}.  The neglect of reflection causes a major discrepancy in the damping and heating rate, as some Poynting flux is expected to propagate back into the corona.}
\label{fig:LC_f1}
\end{figure*}

The damping and heating profiles are shown in Figure \ref{fig:LC_f1}, which can be compared to the weak reflection case shown in Figure \ref{fig:LC_sim}.  As before, blue shows time 3\,s and red 4\,s into the simulation, near the top of the chromosphere shortly after the pulse reaches the transition region.  Between the two times, some of the pulse reflects in the Lare simulation, which reduces the Poynting flux at the leading edge, causing some to begin to propagate in the opposite direction.  This is why the width of the pulse is larger and why the Poynting flux appears to have damped more than in HYDRAD.  The derived heating rates are therefore slightly higher and occur over a smaller height range in HYDRAD than they are in Lare.

The primary differences between the two simulations therefore appear to be due to the neglect of reflection, which would be an important generalization.  While an empirical fit can give an estimate of the transmission coefficient at the transition region, a more general treatment needs to allow for reflection at any strong gradient in the Alfv\'en speed (for example, at shocks).  It is not currently clear how to parameterize the reflection coefficient for any given gradient, so further work is required.

\acknowledgments  The authors thank the anonymous referee for numerous detailed comments that improved the clarity of the paper significantly, and Harry Warren for comments on an early version of the manuscript.  This paper has used color blind friendly color tables, provided by Paul Wright and Graham Kerr (see also \citealt{wright2017}).  This research was performed while JWR held an NRC Research Associateship award at the US Naval Research Laboratory with the support of NASA.  JEL is funded by NASA's HSR and LWS programs.  LAT was supported by the Chief of Naval Research while an NRC Research Associate at the US Naval Research Laboratory.  The authors benefited from participation in the International Space Science Institute team on ``Magnetic Waves in Solar Flares: Beyond the `Standard' Flare Model,'' led by Alex Russell and Lyndsay Fletcher.  This research has made use of NASA's Astrophysics Data System.  CHIANTI is a collaborative project involving George Mason University, the University of Michigan (USA) and the University of Cambridge (UK).  The Lare3D simulations were performed on NASA High End Computing facilities.

\bibliographystyle{apj}

\begin{thebibliography}{}

\bibitem[\protect\citeauthoryear{{Alfv{\'e}n}}{{Alfv{\'e}n}}{1942}]{alfven1942}
{Alfv{\'e}n}, H. 1942, \nat, 150, 405

\bibitem[\protect\citeauthoryear{{Alfv{\'e}n}}{{Alfv{\'e}n}}{1947}]{alfven1947}
{Alfv{\'e}n}, H. 1947, \mnras, 107, 211

\bibitem[\protect\citeauthoryear{{Arber}, {Brady}, \& {Shelyag}}{{Arber}
  et~al.}{2016}]{arber2016}
{Arber}, T.~D., {Brady}, C.~S.,  \& {Shelyag}, S. 2016, \apj, 817, 94

\bibitem[\protect\citeauthoryear{{Arber} et~al.}{{Arber}
  et~al.}{2001}]{arber2001}
{Arber}, T.~D., {Longbottom}, A.~W., {Gerrard}, C.~L.,  \& {Milne}, A.~M. 2001,
  Journal of Computational Physics, 171, 151

\bibitem[\protect\citeauthoryear{{Belcher}}{{Belcher}}{1971}]{belcher1971}
{Belcher}, J.~W. 1971, \apj, 168, 509

\bibitem[\protect\citeauthoryear{{Belcher} \& {Davis}}{{Belcher} \&
  {Davis}}{1971}]{belcherdavis1971}
{Belcher}, J.~W.,  \& {Davis}, L., Jr. 1971, \jgr, 76, 3534

\bibitem[\protect\citeauthoryear{{Birn} et~al.}{{Birn} et~al.}{2009}]{birn2009}
{Birn}, J., {Fletcher}, L., {Hesse}, M.,  \& {Neukirch}, T. 2009, \apj, 695,
  1151

\bibitem[\protect\citeauthoryear{{Bradshaw} \& {Cargill}}{{Bradshaw} \&
  {Cargill}}{2013}]{bradshaw2013}
{Bradshaw}, S.~J.,  \& {Cargill}, P.~J. 2013, \apj, 770, 12

\bibitem[\protect\citeauthoryear{{Bradshaw} \& {Mason}}{{Bradshaw} \&
  {Mason}}{2003}]{bradshaw2003}
{Bradshaw}, S.~J.,  \& {Mason}, H.~E. 2003, \aap, 401, 699

\bibitem[\protect\citeauthoryear{{Bradshaw} \& {Raymond}}{{Bradshaw} \&
  {Raymond}}{2013}]{bradshaw2013b}
{Bradshaw}, S.~J.,  \& {Raymond}, J. 2013, \ssr, 178, 271

\bibitem[\protect\citeauthoryear{{Brady} \& {Arber}}{{Brady} \&
  {Arber}}{2016}]{Brady2016}
{Brady}, C.~S.,  \& {Arber}, T.~D. 2016, \apj, 829, 80

\bibitem[\protect\citeauthoryear{{Cowling}}{{Cowling}}{1956}]{cowling1956}
{Cowling}, T.~G. 1956, \mnras, 116, 114

\bibitem[\protect\citeauthoryear{{de Pontieu}}{{de
  Pontieu}}{1999}]{depontieu1999}
{de Pontieu}, B. 1999, \aap, 347, 696

\bibitem[\protect\citeauthoryear{{De Pontieu}, {Martens}, \& {Hudson}}{{De
  Pontieu} et~al.}{2001}]{depontieu2001}
{De Pontieu}, B., {Martens}, P.~C.~H.,  \& {Hudson}, H.~S. 2001, \apj, 558, 859

\bibitem[\protect\citeauthoryear{{De Pontieu} et~al.}{{De Pontieu}
  et~al.}{2007}]{depontieu2007}
{De Pontieu}, B., et~al. 2007, Science, 318, 1574

\bibitem[\protect\citeauthoryear{{DeForest}}{{DeForest}}{2004}]{deforest2004}
{DeForest}, C.~E. 2004, \apjl, 617, L89

\bibitem[\protect\citeauthoryear{{Del Zanna} et~al.}{{Del Zanna}
  et~al.}{2015}]{delzanna2015}
{Del Zanna}, G., {Dere}, K.~P., {Young}, P.~R., {Landi}, E.,  \& {Mason}, H.~E.
  2015, \aap, 582, A56

\bibitem[\protect\citeauthoryear{{Dere} et~al.}{{Dere} et~al.}{1997}]{dere1997}
{Dere}, K.~P., {Landi}, E., {Mason}, H.~E., {Monsignori Fossi}, B.~C.,  \&
  {Young}, P.~R. 1997, \aaps, 125

\bibitem[\protect\citeauthoryear{{Elmore} et~al.}{{Elmore}
  et~al.}{2014}]{elmore2014}
{Elmore}, D.~F., et~al. 2014, in \procspie, Vol. 9147, Ground-based and
  Airborne Instrumentation for Astronomy V, 914707

\bibitem[\protect\citeauthoryear{{Emslie}}{{Emslie}}{1978}]{emslie1978}
{Emslie}, A.~G. 1978, \apj, 224, 241

\bibitem[\protect\citeauthoryear{{Emslie} \& {Machado}}{{Emslie} \&
  {Machado}}{1979}]{emslie1979}
{Emslie}, A.~G.,  \& {Machado}, M.~E. 1979, \solphys, 64, 129

\bibitem[\protect\citeauthoryear{{Emslie} \& {Sturrock}}{{Emslie} \&
  {Sturrock}}{1982}]{emslie1982}
{Emslie}, A.~G.,  \& {Sturrock}, P.~A. 1982, \solphys, 80, 99

\bibitem[\protect\citeauthoryear{{Fermi}}{{Fermi}}{1949}]{fermi1949}
{Fermi}, E. 1949, Physical Review, 75, 1169

\bibitem[\protect\citeauthoryear{{Fletcher} \& {Hudson}}{{Fletcher} \&
  {Hudson}}{2008}]{fletcher2008}
{Fletcher}, L.,  \& {Hudson}, H.~S. 2008, \apj, 675, 1645

\bibitem[\protect\citeauthoryear{{Goodman}}{{Goodman}}{2011}]{goodman2011}
{Goodman}, M.~L. 2011, \apj, 735, 45

\bibitem[\protect\citeauthoryear{{Goodman} \& {Kazeminezhad}}{{Goodman} \&
  {Kazeminezhad}}{2010}]{Goodman2010}
{Goodman}, M.~L.,  \& {Kazeminezhad}, F. 2010, \apj, 708, 268

\bibitem[\protect\citeauthoryear{{Haerendel}}{{Haerendel}}{2006}]{haerendel2006}
{Haerendel}, G. 2006, \ssr, 124, 317

\bibitem[\protect\citeauthoryear{{Haerendel}}{{Haerendel}}{2009}]{haerendel2009}
{Haerendel}, G. 2009, \apj, 707, 903

\bibitem[\protect\citeauthoryear{{Haerendel}}{{Haerendel}}{2012}]{haerendel2012}
{Haerendel}, G. 2012, \apj, 749, 166

\bibitem[\protect\citeauthoryear{{Jel{\'{\i}}nek} et~al.}{{Jel{\'{\i}}nek}
  et~al.}{2017}]{jelinek2017}
{Jel{\'{\i}}nek}, P., {Karlick{\'y}}, M., {Van Doorsselaere}, T.,  \&
  {B{\'a}rta}, M. 2017, \apj, 847, 98

\bibitem[\protect\citeauthoryear{{Johnston} et~al.}{{Johnston}
  et~al.}{2017a}]{johnston2017a}
{Johnston}, C.~D., {Hood}, A.~W., {Cargill}, P.~J.,  \& {De Moortel}, I. 2017a,
  \aap, 597, A81

\bibitem[\protect\citeauthoryear{{Johnston} et~al.}{{Johnston}
  et~al.}{2017b}]{johnston2017b}
{Johnston}, C.~D., {Hood}, A.~W., {Cargill}, P.~J.,  \& {De Moortel}, I. 2017b,
  \aap, 605, A8

\bibitem[\protect\citeauthoryear{{Kaufmann} et~al.}{{Kaufmann}
  et~al.}{1984}]{kaufmann1984}
{Kaufmann}, P., {Correia}, E., {Costa}, J.~E.~R., {Dennis}, B.~R., {Hurford},
  G.~J.,  \& {Brown}, J.~C. 1984, \solphys, 91, 359

\bibitem[\protect\citeauthoryear{{Kerr} et~al.}{{Kerr} et~al.}{2016}]{kerr2016}
{Kerr}, G.~S., {Fletcher}, L., {Russell}, A.~J.~B.,  \& {Allred}, J.~C. 2016,
  \apj, 827, 101

\bibitem[\protect\citeauthoryear{{Khomenko}}{{Khomenko}}{2017}]{khomenko2017}
{Khomenko}, E. 2017, Plasma Physics and Controlled Fusion, 59, 014038

\bibitem[\protect\citeauthoryear{{Kigure} et~al.}{{Kigure}
  et~al.}{2010}]{kigure2010}
{Kigure}, H., {Takahashi}, K., {Shibata}, K., {Yokoyama}, T.,  \& {Nozawa}, S.
  2010, \pasj, 62, 993

\bibitem[\protect\citeauthoryear{{Kiplinger} et~al.}{{Kiplinger}
  et~al.}{1983}]{kiplinger1983}
{Kiplinger}, A.~L., {Dennis}, B.~R., {Frost}, K.~J., {Orwig}, L.~E.,  \&
  {Emslie}, A.~G. 1983, \apjl, 265, L99

\bibitem[\protect\citeauthoryear{{Klimchuk}}{{Klimchuk}}{2006}]{klimchuk2006}
{Klimchuk}, J.~A. 2006, \solphys, 234, 41

\bibitem[\protect\citeauthoryear{{Laming}}{{Laming}}{2015}]{laming2015}
{Laming}, J.~M. 2015, Living Reviews in Solar Physics, 12, 2

\bibitem[\protect\citeauthoryear{{Lazarian}}{{Lazarian}}{2016}]{lazarian2016}
{Lazarian}, A. 2016, \apj, 833, 131

\bibitem[\protect\citeauthoryear{{Leake}, {Arber}, \& {Khodachenko}}{{Leake}
  et~al.}{2005}]{leake2005}
{Leake}, J.~E., {Arber}, T.~D.,  \& {Khodachenko}, M.~L. 2005, \aap, 442, 1091

\bibitem[\protect\citeauthoryear{{Leake} et~al.}{{Leake}
  et~al.}{2014}]{leake2014}
{Leake}, J.~E., et~al. 2014, \ssr, 184, 107

\bibitem[\protect\citeauthoryear{{Leake} \& {Linton}}{{Leake} \&
  {Linton}}{2013}]{Leake2013}
{Leake}, J.~E.,  \& {Linton}, M.~G. 2013, \apj, 764, 54

\bibitem[\protect\citeauthoryear{{Liu} et~al.}{{Liu} et~al.}{2016}]{liu2016}
{Liu}, C., et~al. 2016, Nature Communications, 7, 13104

\bibitem[\protect\citeauthoryear{{Lundquist}}{{Lundquist}}{1949a}]{lundquist1949a}
{Lundquist}, S. 1949a, \nat, 164, 145

\bibitem[\protect\citeauthoryear{{Lundquist}}{{Lundquist}}{1949b}]{lundquist1949b}
{Lundquist}, S. 1949b, Physical Review, 76, 1805

\bibitem[\protect\citeauthoryear{{Machado}, {Emslie}, \& {Brown}}{{Machado}
  et~al.}{1978}]{machado1978}
{Machado}, M.~E., {Emslie}, A.~G.,  \& {Brown}, J.~C. 1978, \solphys, 58, 363

\bibitem[\protect\citeauthoryear{{Mart{\'{\i}}nez-Sykora}
  et~al.}{{Mart{\'{\i}}nez-Sykora} et~al.}{2017}]{martinezsykora2017}
{Mart{\'{\i}}nez-Sykora}, J., {De Pontieu}, B., {Carlsson}, M., {Hansteen},
  V.~H., {N{\'o}brega-Siverio}, D.,  \& {Gudiksen}, B.~V. 2017, \apj, 847, 36

\bibitem[\protect\citeauthoryear{{Mart{\'{\i}}nez-Sykora}, {De Pontieu}, \&
  {Hansteen}}{{Mart{\'{\i}}nez-Sykora} et~al.}{2012}]{martinezsykora2012}
{Mart{\'{\i}}nez-Sykora}, J., {De Pontieu}, B.,  \& {Hansteen}, V. 2012, \apj,
  753, 161

\bibitem[\protect\citeauthoryear{{McIntosh} et~al.}{{McIntosh}
  et~al.}{2011}]{mcintosh2011}
{McIntosh}, S.~W., {de Pontieu}, B., {Carlsson}, M., {Hansteen}, V., {Boerner},
  P.,  \& {Goossens}, M. 2011, \nat, 475, 477

\bibitem[\protect\citeauthoryear{{Ni} et~al.}{{Ni} et~al.}{2015}]{Ni2015}
{Ni}, L., {Kliem}, B., {Lin}, J.,  \& {Wu}, N. 2015, \apj, 799, 79

\bibitem[\protect\citeauthoryear{{Oliver} et~al.}{{Oliver}
  et~al.}{1993}]{oliver1993}
{Oliver}, R., {Ballester}, J.~L., {Hood}, A.~W.,  \& {Priest}, E.~R. 1993,
  \aap, 273, 647

\bibitem[\protect\citeauthoryear{{Osterbrock}}{{Osterbrock}}{1961}]{osterbrock1961}
{Osterbrock}, D.~E. 1961, \apj, 134, 347

\bibitem[\protect\citeauthoryear{{Parker}}{{Parker}}{1991}]{parker1991}
{Parker}, E.~N. 1991, \apj, 372, 719

\bibitem[\protect\citeauthoryear{{Pasachoff} et~al.}{{Pasachoff}
  et~al.}{2002}]{pasachoff2002}
{Pasachoff}, J.~M., {Babcock}, B.~A., {Russell}, K.~D.,  \& {Seaton}, D.~B.
  2002, \solphys, 207, 241

\bibitem[\protect\citeauthoryear{{Piddington}}{{Piddington}}{1956}]{piddington1956}
{Piddington}, J.~H. 1956, \mnras, 116, 314

\bibitem[\protect\citeauthoryear{{Reale}}{{Reale}}{2014}]{reale2014}
{Reale}, F. 2014, Living Reviews in Solar Physics, 11, 4

\bibitem[\protect\citeauthoryear{{Reep}, {Bradshaw}, \& {Alexander}}{{Reep}
  et~al.}{2015}]{reep2015}
{Reep}, J.~W., {Bradshaw}, S.~J.,  \& {Alexander}, D. 2015, \apj, 808, 177

\bibitem[\protect\citeauthoryear{{Reep} \& {Russell}}{{Reep} \&
  {Russell}}{2016}]{reep2016}
{Reep}, J.~W.,  \& {Russell}, A.~J.~B. 2016, \apjl, 818, L20

\bibitem[\protect\citeauthoryear{{Rubio da Costa} et~al.}{{Rubio da Costa}
  et~al.}{2016}]{rubiodacosta2016}
{Rubio da Costa}, F., {Kleint}, L., {Petrosian}, V., {Liu}, W.,  \& {Allred},
  J.~C. 2016, \apj, 827, 38

\bibitem[\protect\citeauthoryear{{Rudawy} et~al.}{{Rudawy}
  et~al.}{2010}]{rudawy2010}
{Rudawy}, P., {Phillips}, K.~J.~H., {Buczylko}, A., {Williams}, D.~R.,  \&
  {Keenan}, F.~P. 2010, \solphys, 267, 305

\bibitem[\protect\citeauthoryear{{Rudawy} et~al.}{{Rudawy}
  et~al.}{2004}]{rudawy2004}
{Rudawy}, P., {Phillips}, K.~J.~H., {Gallagher}, P.~T., {Williams}, D.~R.,
  {Rompolt}, B.,  \& {Keenan}, F.~P. 2004, \aap, 416, 1179

\bibitem[\protect\citeauthoryear{{Russell}}{{Russell}}{2017}]{russell2017}
{Russell}, A.~J.~B. 2017, ArXiv e-prints

\bibitem[\protect\citeauthoryear{{Russell} \& {Fletcher}}{{Russell} \&
  {Fletcher}}{2013}]{russellfletcher2013}
{Russell}, A.~J.~B.,  \& {Fletcher}, L. 2013, \apj, 765, 81

\bibitem[\protect\citeauthoryear{{Russell} et~al.}{{Russell}
  et~al.}{2016}]{russell2016}
{Russell}, A.~J.~B., {Mooney}, M.~K., {Leake}, J.~E.,  \& {Hudson}, H.~S. 2016,
  \apj, 831, 42

\bibitem[\protect\citeauthoryear{{Russell} \& {Stackhouse}}{{Russell} \&
  {Stackhouse}}{2013}]{russellstackhouse2013}
{Russell}, A.~J.~B.,  \& {Stackhouse}, D.~J. 2013, \aap, 558, A76

\bibitem[\protect\citeauthoryear{{Shibata} \& {Moriyasu}}{{Shibata} \&
  {Moriyasu}}{2003}]{shibata2003}
{Shibata}, K.,  \& {Moriyasu}, S. 2003, in Astronomical Society of the Pacific
  Conference Series, Vol. 286, Current Theoretical Models and Future High
  Resolution Solar Observations: Preparing for ATST, ed. A.~A. {Pevtsov} \&
  H.~{Uitenbroek}, 377

\bibitem[\protect\citeauthoryear{{Soler}, {Ballester}, \&
  {Zaqarashvili}}{{Soler} et~al.}{2015}]{soler2015a}
{Soler}, R., {Ballester}, J.~L.,  \& {Zaqarashvili}, T.~V. 2015, \aap, 573, A79

\bibitem[\protect\citeauthoryear{{Soler}, {Carbonell}, \& {Ballester}}{{Soler}
  et~al.}{2015}]{soler2015b}
{Soler}, R., {Carbonell}, M.,  \& {Ballester}, J.~L. 2015, \apj, 810, 146

\bibitem[\protect\citeauthoryear{{Soler} et~al.}{{Soler}
  et~al.}{2013}]{soler2013}
{Soler}, R., {Carbonell}, M., {Ballester}, J.~L.,  \& {Terradas}, J. 2013,
  \apj, 767, 171

\bibitem[\protect\citeauthoryear{{Soler} et~al.}{{Soler}
  et~al.}{2016}]{Soler2016}
{Soler}, R., {Terradas}, J., {Oliver}, R.,  \& {Ballester}, J.~L. 2016, \aap,
  592, A28

\bibitem[\protect\citeauthoryear{{Suzuki} \& {Inutsuka}}{{Suzuki} \&
  {Inutsuka}}{2005}]{suzuki2005}
{Suzuki}, T.~K.,  \& {Inutsuka}, S.-i. 2005, \apjl, 632, L49

\bibitem[\protect\citeauthoryear{{Takeuchi} \& {Shibata}}{{Takeuchi} \&
  {Shibata}}{2001}]{takeuchi2001}
{Takeuchi}, A.,  \& {Shibata}, K. 2001, \apjl, 546, L73

\bibitem[\protect\citeauthoryear{{Tarr}}{{Tarr}}{2017}]{tarr2017}
{Tarr}, L.~A. 2017, \apj, 847, 1

\bibitem[\protect\citeauthoryear{{Tomczyk} et~al.}{{Tomczyk}
  et~al.}{2007}]{tomczyk2007}
{Tomczyk}, S., {McIntosh}, S.~W., {Keil}, S.~L., {Judge}, P.~G., {Schad}, T.,
  {Seeley}, D.~H.,  \& {Edmondson}, J. 2007, Science, 317, 1192

\bibitem[\protect\citeauthoryear{{van Ballegooijen} \& {Asgari-Targhi}}{{van
  Ballegooijen} \& {Asgari-Targhi}}{2016}]{vanballegooijen2016}
{van Ballegooijen}, A.~A.,  \& {Asgari-Targhi}, M. 2016, \apj, 821, 106

\bibitem[\protect\citeauthoryear{{van Ballegooijen} et~al.}{{van Ballegooijen}
  et~al.}{2011}]{vanballegooijen2011}
{van Ballegooijen}, A.~A., {Asgari-Targhi}, M., {Cranmer}, S.~R.,  \& {DeLuca},
  E.~E. 2011, \apj, 736, 3

\bibitem[\protect\citeauthoryear{Wright}{Wright}{2017}]{wright2017}
Wright, P.~J. 2017, {ColourBlind: A Collection of IDL Colour-blind-friendly
  Colour Tables}

\bibitem[\protect\citeauthoryear{{Zweibel}}{{Zweibel}}{1989}]{zweibel1989}
{Zweibel}, E.~G. 1989, \apj, 340, 550

\bibitem[\protect\citeauthoryear{{Zweibel} et~al.}{{Zweibel}
  et~al.}{2011}]{zweibel2011}
{Zweibel}, E.~G., {Lawrence}, E., {Yoo}, J., {Ji}, H., {Yamada}, M.,  \&
  {Malyshkin}, L.~M. 2011, Physics of Plasmas, 18, 111211

\end{thebibliography}

\end{document}